\documentclass[10pt,aps,pre,onecolumn,superscriptaddress,floatfix,nofootinbib,notitlepage,preprintnumbers]{revtex4-1}
\usepackage{amsmath,amssymb,amsfonts}
\usepackage{graphicx,color}
\usepackage{dcolumn}
\usepackage{bm}
\usepackage{mathtools}

\begin{document}

\title{Entangling credit and funding shocks in interbank markets}

\author{Giulio Cimini}
\affiliation{\small IMT - School for Advanced Studies, 55100 Lucca (Italy)}
\affiliation{\small Istituto dei Sistemi Complessi (ISC-CNR), 00185 Rome (Italy)}
\author{Matteo Serri}
\affiliation{\small Universit\`a ``Sapienza'', Facolt\`a di Economia, Dipartimento MEMOTEF, 00185 Rome (Italy)}

\begin{abstract}
Credit and liquidity risks represent main channels of financial contagion for interbank lending markets. 
On one hand, banks face potential losses whenever their counterparties are under distress and thus unable to fulfill their obligations. 
On the other hand, solvency constraints may force banks to recover lost fundings by selling their illiquid assets, 
resulting in effective losses in the presence of fire sales---that is, when funding shortcomings are widespread over the market. 
Because of the complex structure of the network of interbank exposures, these losses reverberate among banks and eventually get amplified, 
with potentially catastrophic consequences for the whole financial system. Building on Debt Rank~\cite{Battiston2012}, 
in this work we define a systemic risk metric that estimates the potential amplification of losses 
in interbank markets accounting for both credit and liquidity contagion channels: the Debt-Solvency Rank. 
We implement this framework on a dataset of 183 European banks that were publicly traded between 2004 and 2013, 
showing indeed that liquidity spillovers substantially increase systemic risk, and thus cannot be neglected in stress-test scenarios. 
We also provide additional evidence that the interbank market was extremely fragile up to the 2008 financial crisis, becoming slightly more robust only afterwards.
\end{abstract} 

\maketitle

\section*{Introduction}

The recent financial crises have highlighted the importance to properly assess systemic risk in capital markets. 
In particular, both researchers and regulators realized that the financial system is actually more fragile than previously thought, 
because of the complexity of interconnections between financial institutions~\cite{Lau2009,Brunnermeier2009a,Fouque2013,Battiston2015price,Battiston2016}, 
resulting both from direct exposures to bilateral contracts and from indirect exposures to common assets \cite{Allen2000,Gai2011,Bluhm2011,Acemoglu2015,Greenwood2015}. 
Indeed, while interconnectedness means diversification and thus reduces individual risk, it can however increase systemic risk: 
financial distress can spread between institutions through such exposures and propagate over the market, leading to amplification effects like default cascades \cite{Corsi2013,Caccioli2014,Bardoscia2016X}. 
Thus, much attention is being devoted nowadays to characterize the emerging network structure of financial markets from the viewpoint of systemic risk~\cite{Boss2004,Iori2006,Elsinger2006,Nier2008}---especially 
in terms of counterparty and rollover risks caused by loans between institutions~\cite{Gai2010,Haldane2011}. 
Remarkably, the network-based approach allows to track the reverberation of a credit event or liquidity squeeze throughout the financial system, 
with important outputs like measures for potential capital losses and domino effect, and the identification of systemic and vulnerable institutions~\cite{Lau2009}. 
The shortcoming of requiring data on individual exposures, which are often privacy-protected and thus unaccessible, 
can be overcome by using either stylized banking systems (see, e.g.,~\cite{Nier2008,Krause2012}) or effective methods to reconstruct the network structure from the available data~\cite{Caldarelli2013} 
(usually, the balance sheet composition of financial institution)~\cite{Wells2004,Upper2011,Mastromatteo2012,Baral2012,Drehmann2013,Halaj2013,Anand2014,Montagna2014,Peltonen2015,Cimini2015,Cimini2015b}.

Here we focus on the interbank lending market, where banks temporarily short on liquidity borrow money for a specified term (mostly overnight) from other banks having excess liquidity, 
that in turn receive an interest on the loan. This market plays crucial role in the financial system, by allowing banks to cope with liquidity fluctuations 
and meet reserve requirements at the end of the trading day~\cite{Wiemers2003,Angelini2011,Allen2014}. 
Because of its structure, the interbank market can be properly represented as a directed weighted network, where interbank loans constitute the direct exposures between banks 
and allow for the propagation of financial distress in the system through two main mechanisms~\cite{Nier2008,Lau2009,Krause2012}: {\em credit shocks} and {\em funding shocks}. 
Credit losses are related to counterparty risk, and are faced by lender banks when their borrower banks default and fail to fulfill their obligations.  
These losses can then lead to the default of lenders, resulting in another wave of credit shocks. In a seminal paper, Eisenberg and Noe~\cite{Eisenberg2001} developed an analytical framework 
to determine the set of payments that clear the network given an initial shock, assessing in this way how losses propagate through the system (see also~\cite{Elsinger2006,Glasserman2015}). 
The Furfine algorithm~\cite{Furfine2003} instead quantifies losses due to financial contagion by iteratively propagating distress from a defaulted bank to its creditors. 
In contrast, in the Debt Rank algorithm~\cite{Battiston2012} credit shocks propagate also in the absence of defaults, provided that balance sheets are deteriorated: 
{\em potential} losses in the equity of a borrower translate into the devaluation of interbank assets of the corresponding lender~\cite{Bardoscia2015,Bardoscia2015X}.
In this way Debt Rank accounts for the incremental build-up of distress in the system even before the occurrence of defaults, and thus seems appropriate for stress-test scenarios~\cite{Battiston2015X}.

Funding shocks are instead related to rollover risk, and concern financial institutions subject to regulatory solvency constraints marking their assets to market 
in order to replace lost fundings~\cite{Cifuentes2005,Lau2009,Kapadia2012}. If the interbank market is under distress and shrinks, banks short on liquidity 
may be unable to borrow all the money they need from the market, and be forced to sell their illiquid assets~\cite{Brunnermeier2009b}. 
In the presence of fire sales, the market demand for illiquid assets becomes inelastic, depressing the market prices of these assets and resulting in effective losses for banks~\cite{Adrian2010,Diamond2010}. 
These fire sales spillovers create the incentive to hoard liquidity~\cite{Berrospide2012,Gale2013}, which can in turn induce another wave of sales, activating a liquidity spiral 
that can lead the interbank market to freeze completely~\cite{Acharya2011,Acharya2013,Gabrieli2014}. Note that beyond a direct channel (funding losses combined with cash-flow constraints), 
fire sales may also originate indirectly through common assets holdings, or the leverage targeting policy adopted by banks~\cite{Greenwood2015,Duarte2015,Battiston2015X}. 
In any event, liquidity spillovers were shown to represent an important dimension of overall systemic risk, comparable to counterparty domino effects~\cite{Glasserman2015}.

In this work we aim at combining these two channels of financial contagion into a systemic risk metric for the spreading of potential losses in interbank markets. 
By doing so, we define the Debt-Solvency Rank, an extension of the Debt Rank framework to such a credit-funding shock scenario~\cite{Lau2009}. We build on the assumption 
that equity losses experienced by a bank do imply not only a decreasing value of its obligations, but also a decreasing ability to lend money to the market---even if no default has occurred. 
Such financial distress then reverberates throughout the market, turning into equity losses for other banks. We show that accounting for fire sales spillovers directly within the distress propagation dynamics 
leads to a more refined assessment of systemic risk for interbank markets, where liquidity issues represent a first-order correction to counterparty risk. 
We apply our method on a dataset of 183 European banks from 2004 to 2013, quantifying individual impact and vulnerability of these financial institutions over time. 
Our analysis confirms that liquidity issues are important for a correct systemic risk assessment, their effect being particularly relevant when credit risk peaks. 
We also provide additional evidence that the interbank market was extremely fragile up to the 2008 financial crisis, becoming slightly more robust only afterwards.

\section*{Data \& Methods}

\subsection*{Balance sheets basics and the interbank network}

In order to characterize the interbank market, we use the Bureau van Dijk Bankscope database\footnote{Raw data are available from Bureau van Dijk: https://bankscope.bvdinfo.com. 
Refer to~\cite{Battiston2015X} for all the details about the handling of missing data.} that contains yearly-aggregated balance sheets information of $N=183$ large European banks, from 2004 to 2013. 
For each bank $i$ and for each year $T$ (we omit the year label for convenience), the balance sheet composition summarizes the financial position of the bank, 
and consists of assets with a positive economic value (loans, derivatives, stocks, bonds, real estate, and so on) and of liabilities with a negative economic value (such as customer deposits, and debits). 
To disentangle the interbank market from the overall financial system, we distinguish between interbank assets $A_i$ (corresponding to the total loans to other banks) 
and external (non-interbank) assets $\tilde{A}_i$, and between interbank liabilities $L_i$ (the total debts to other banks) and external (non-interbank) liabilities $\tilde{L}_i$. 
The balance sheet identity for bank $i$ defines its equity $E_i$ (or tier capital) as the difference between its total assets and liabilities: 
\begin{equation}\label{eq:balancesheet} 
E_i=(\tilde{A}_i+A_i)-(\tilde{L}_i+L_i).
\end{equation}
A bank is solvent as long as its equity is positive. In fact, negative equity means insolvency, as the bank cannot pay back its debs even by selling all of its assets. 
Here, following the literature on financial contagion~\cite{Nier2008,Gai2010,Upper2011}, we take insolvency as a proxy for default. 
In the following, we denote the overall equity of the interbank market as $E=\sum_iE_i$, and its total volume as $C=\sum_iA_i\equiv\sum_iL_i$ (see Table \ref{tab:stats}). 

For each bank $i$, its interbank assets and liabilities are aggregates of the individual loans to and borrowings from other banks: 
$A_i=\sum_j A_{ij}$ and $L_i=\sum_j L_{ij}$, where $A_{ij}$ is the amount of the loan extended by bank $i$ to bank $j$, 
which represents an asset for bank $i$ and a liability for bank $j$: thus $L_{ji}\equiv A_{ij}$ $\forall i,j$. 
Hence, the interbank market can be represented as a directed weighted network, in which links correspond to exposures between banks~\cite{Iori2008,Barucca2016}. 
However, because of privacy issues, we do not have information on the individual loans and obligations. Yet, we can reconstruct the interbank network of exposures 
from aggregated interbank assets and liabilities by resorting to the two-step inference procedure described in~\cite{Cimini2015}\footnote{We set a network density of 10\% 
and generate an ensemble of 1000 interbank networks. Results presented in this paper are averaged over this ensemble.}, which was shown to be particularly effective 
in assessing the presence and extent of overnight interbank loans. In a nutshell, the method estimates individual exposures as:
\begin{equation}\label{eq:weight}
A_{ij} =\frac{z^{-1}+A_i\,L_j}{C}\,a_{ij},\qquad a_{ij}=
\begin{cases}
1\quad\mbox{with probability }p_{ij}=(A_i\,L_j)/(z^{-1}+\,A_i\,L_j)\\
0\quad\mbox{otherwise}
\end{cases}
\end{equation}
where $z$ is a parameter that controls for the density of the network and $a_{ij}$ denotes the presence of the link.
Note that the probabilistic nature of the connections is particularly fit to represent overnight positions: 
we assume that contracts are continuously placed and immediately resolved and rolled over, 
a scenario in which credit and liquidity shocks can be assumed to hit the system and spread on the same temporal scale.

\begin{table}[t!]
\centering
\begin{tabular}{l|r|r|r|r|r|r|r|r|r|r}
year 	&	2004	&	2005	&	2006	&	2007	&	2008	&	2009	&	2010	&	2011	&	2012	&	2013	\\
\hline
total equity ($E$) 	&	 496976 	&	 900950 	&	 1207734 	&	 1542098 	&	 1291499 	&	 1680088 	&	 1708205 	&	 1629743 	&	 1699175 	&	 1778428 	\\
total volume ($C$) 	&	 1424469 	&	 2453230 	&	 3063762 	&	 3874003 	&	 3040012 	&	 2728253 	&	 2371510 	&	 2286400 	&	 2137298 	&	 2008040 	\\
\end{tabular}
\caption{Aggregate equity and volume of the interbank market (expressed in million USD) for the various years considered.}
\label{tab:stats}
\end{table}

\subsection*{Credit \& funding shocks}

Having built the network of interbank exposures, we now describe the credit-funding shock dynamics on the interbank market~\cite{Lau2009,Krause2012} that will be at the basis of the Debt-Solvency Rank. 
Suppose a generic bank $u$ is hit by an exogenous shock (such as the sudden devaluation of its external assets) that causes $u$ to become insolvent. 
This event triggers a cascade of losses in the network, through two distinct processes.
 \begin{enumerate}
  \item {\em Credit shock}: bank $u$ fails to meet its obligations, resulting in effective losses for its creditors. 
  In particular, each other bank $j$ suffers a loss equal to $\lambda A_{ju}$, where $0\le\lambda\le1$ indicates the amount of loss given default. 
  Here we consider only uncollateralized markets (i.e., without a central counterparty that guarantees for interbank loans), and thus set $\lambda=1$ always\footnote{In the literature, 
  $\lambda$ is usually set in one of two ways: i) exogenously to a constant value, $\lambda=0.4\div0.45$ \cite{Kaufman1994}; 
  ii) stochastically from a distribution $q(\lambda)$ \cite{Memmel2012}. The second approach reflects the fact that $\lambda$ usually varies considerably, 
  with a large portion of the probability mass at 0\% (collateralized loans) and at 100\% (completely unsecured loans). 
  Empirical studies \cite{Memmel2012} suggest to use a beta distribution $q(\lambda)=\lambda^{\alpha-1}(1-\lambda)^{\beta-1}\,\Gamma(\alpha+\beta)/[\Gamma(\alpha)\Gamma(\beta)]$, 
  with shape parameters $\alpha=0.28$ and $\beta=0.35$.} 
  \item {\em Funding shock}: banks are unable to replace all the liquidity previously granted by $u$; this, in turn, triggers a fire sale of assets. In particular, 
  each other bank $j$ can use its cash reserves to replenish only a fraction $(1-\rho)$ of the lost funding from $u$, and its illiquid assets trade at a discount (their market value is less that their book value), 
  so that $j$ must sell assets worth $(1+\gamma)\rho A_{uj}$ in book value terms---corresponding to an overall loss of $\gamma\rho A_{uj}$, where the parameter $\gamma$ set the change in asset prices.
  In the following we will focus on two cases: $\rho=1$, meaning that banks actually cannot replace the lost funding from $u$ and are thus forced to entirely replenish the corresponding liquidity by assets sales; 
  and $\rho=0$, for which there are no funding shocks as banks always have enough liquid assets to cope with these losses.
 \end{enumerate}
Overall, the generic bank $j$ faces a loss of $\lambda A_{ju}+\gamma\rho A_{uj}$ (the first term appears if $j$ is a lender of $u$, and the second one if $j$ is a borrower of $u$).  
Then, bank $j$ also defaults if this loss exceeds $E_j$, and a new credit-funding shock propagates through the market, resulting in a wave of cascade failures. 

\medskip
Before moving further, we discuss how to compute the coefficient $\gamma$ that determines the losses due to funding shocks. 
Following a common approach \cite{Nier2008,Ellul2011,Feldhutter2012,Greenwood2015}, here we assume that assets fire sales generate a linear impact on prices: 
given that $Q=\rho\sum_j A_{uj}$ is the aggregate amount of assets that need to be liquidated, we set the price impact coefficient such that when $Q=0$ the relative assets price change is $\Delta p/p=0$ 
(i.e., no devaluation), and when $Q=C$ it is $\Delta p/p=-1$ (assets value vanishes). This results in a price impact $\Delta p/p=-Q/C$.\footnote{Note that the coefficient $C^{-1}$ in our case 
is of order $10^{-13}$ (see Table \ref{tab:stats}), meaning that 10bn euros of trading imbalances lead to a price change of order ten basis points---in line with \cite{Greenwood2015}.} 
Finally, to obtain the corresponding value of $\gamma$, it is sufficient to equate the relative loss $\gamma\rho$ to the relative amount sold $(1+\gamma)\rho$ times the relative assets price change $(-\Delta p/p)$, 
obtaining $\gamma=[C/Q-1]^{-1}$.

\subsection*{The Debt-Solvency (DS) Rank}

We are now equipped to apply the credit-funding shock scenario described above to the Debt Rank framework. Before starting, we recall again the basic idea behind Debt Rank: 
the probabilities of banks default can be obtained by iteratively spreading the individual banks distress levels weighted by the potential wealth affected~\cite{Battiston2012}. 
In its original formulation, Debt Rank considers counterparty risk in a network of long-term interbank loans (in physics terminology, the network is said to be {\em quenched}). 
In this situation, liabilities stay at their face value, while assets are marked to market. The microscopic dynamics of distress propagation then works as follows~\cite{Bardoscia2015}. 
If a bank $u$ is hit by a shock and its equity decreases, also if $u$ is still solvent the market value of its obligations decreases, because $u$ is now ``closer'' to default 
and thus less likely to meet its obligation at maturity. This results in a loss on equity for the banks that have $u$'s obligations in their balance sheet, which in turn decrease the market value 
of these banks obligations, and so forth. Now, the Debt Rank assumes that relative changes of equity translate linearly into relative changes of asset values, resulting in an {\em impact} 
of bank $u$ on bank $j$ equal to $\lambda\Lambda_{ju}=\lambda A_{ju}/E_j$, where $\mathbf{\Lambda}$ is the interbank leverage matrix~\cite{Bardoscia2015,Battiston2015X}.\footnote{The same approach 
can be extended to non-linear impact functions~\cite{Bardoscia2015X}} Then, by iterating the described credit shocks dynamics, Debt Rank allows to compute the final level of financial distress in the network. 

In contrast, in the interbank market the majority of contracts have overnight duration, meaning that they are placed and shortly after resolved and rolled-over: 
the network is said to be {\em annealed}, as links change continuously. In this situation, liabilities also change and may lead to funding shocks. 
In line with the original Debt Rank approach, we can assume that the ability of bank $u$ to lend money to the market decreases proportionally to its equity, 
and that the liabilities (borrowings) of other banks towards $u$ change in the same way. As funding shocks results in effective equity losses only in the presence of fire sales 
(i.e., when $\gamma>0$), the impact of bank $u$ on bank $j$ in this case can be assessed as $\rho\gamma\Upsilon_{ju}=\rho\gamma A_{uj}/E_j$. 
Overall, the impact of bank $u$ on bank $j$ due to both counterparty creditworthiness and liquidity shortage issues is $\lambda\Lambda_{ju}+\rho\gamma\Upsilon_{ju}$. 
Importantly, we assume that these shocks happen on the same temporal scale. In this way, we can extend the Debt Rank formulation to a dynamics where the equity decrease of bank $u$ 
causes an equity decrease both for the lenders of $u$ because of credit shocks, and at the same time for the borrowers of $u$ because of funding shocks. 

We can formalize the above discussion as follows. The dynamics of shock propagation consists of several rounds $\{t\}$, and we are interested in quantifying the level of financial distress 
of each bank $i$ and at each time step $t$, given by the relative change of equity: $h_i(t)=1-E_i(t)/E_i(0)$. 
By definition, $h_i=0$ when no equity losses occurred for bank $i$, $h_i=1$ when that bank defaults, and $0<h_i<1$ for intermediate distress levels. 
At step $t=0$, there is no distress in the system, and thus $h_i(0)=0$ $\forall i$. At step $t=1$, an exogenous shock hits the market causing a decrease of equity for some banks; 
hence, the set $\{h_i(1)\}$ represents the initial condition of the dynamics. Subsequent values of $h$ are obtained by spreading this shock on the system, 
and writing up the equation for the evolution of banks equity~\cite{Bardoscia2015}. 
By defining $\mathcal{A}(t)=\{j:h_j(t-1)<1\}$ as the set of banks that have not defaulted up to time $t$ (and thus can still spread their financial distress), 
we assume that a generic bank $j$ propagates shocks as long as it keeps receiving them, i.e., provided that $h_j(t)>h_j(t-1)$. 
However, such a propagation is damped with a function $D[(t-t_j)/\tau]$, where $t_j$ denotes the time step when bank $j$ first becomes distressed, 
$t_j$ : $h_j(t_j)>0$ and $h_j(t_j-1)=0$, and $\tau$ is the damping scale. Overall, we get:
\begin{equation}\label{eq:acca}
h_i(t+1)=\min\left\{1,\; h_i(t)+\sum_{j\in\mathcal{A}(t)}[\lambda\Lambda_{ij}+\rho\gamma(t)\Upsilon_{ij}]\,[h_j(t)-h_j(t-1)]\,D[(t-t_j)/\tau]\right\}. 
\end{equation}
In this expression, according to the discussion in the previous paragraphs, the individual impact of bank $j$ on bank $i$ at step $t$ is 
$\lambda\Lambda_{ij}+\rho\gamma(t)\Upsilon_{ij}$, where the fire sale devaluation factor depends on time as $\gamma(t)=\{C/[\rho Q(t)]-1\}^{-1}$. 
The aggregate amount of interbank assets potentially to be liquidated at $t$ is in turn given by $Q(t)=\sum_{j\in\mathcal{A}(t)}\sum_k A_{jk}[h_j(t)-h_j(t-1)]\,D[(t-t_j)/\tau]$. 
Concerning the damping function, note that for generic bank $j$ it plays a role only when $t\ge t_j$, as otherwise $h_j(t)=h_j(t-1)=0$. 
Importantly, it must have the properties $D(0)=1$ and $D(x)\xrightarrow{\scriptscriptstyle x\to\infty}0$: the spreading is damped more and more with the propagation step, and eventually vanishes. 
An exponential shape $D[(t-t_j)/\tau]=\exp[-(t-t_j)/\tau]$ satisfies these properties, but alternative forms can be used as well. 
The damping scale $\tau$ sets the mean lifetime of the spreading. In the limit $\tau\to0$, we have $\exp[-(t-t_j)/\tau]\to\delta(t-t_j)$: 
banks spread their distress only once, as in the original version of Debt Rank~\cite{Battiston2012}. In the opposite limit $\tau\to\infty$, $\exp[-(t-t_j)/\tau]\to1$: 
banks always propagate received shocks until they default, as in the most recent version of Debt Rank~\cite{Bardoscia2015}. 

The above described dynamics stops at $t^*$ when no more banks can propagate their distress, i.e., $h_i(t^*)=h_i(t^*-1)$ $\forall i$. 
The overall Debt-Solvency ($DS$) Rank of the network is then obtained as:
\begin{equation}\label{eq:dsrank}
DS(t^*)=\sum_i[h_i(t^*)-h_i(1)]\nu_i\equiv\frac{E(1)-E(t^*)}{E(0)},
\end{equation}
where $\nu_i=E_i(0)/\sum_jE_j(0)=E_i(0)/E(0)$ is the relative importance of bank $i$ in the system, and $E(t)=\sum_iE_i(t)$ is the total equity of the system at time $t$. 
$DS(t^*)$ thus represents the amount of equity that is potentially at risk in the system, given an initial shock $\{h_i(1)\}$. 
Note that the original Debt Rank accounting only for credit shocks is recovered for $\lambda=1$ and $\rho=0$.

\section*{Results}

\subsection*{Group DS Rank}

We first consider the simplest case of an initial shock to the system corresponding to a fixed percentage $(1-\psi)$ of equity loss for each bank: 
$E_i(1)=(1-\psi)E_i(0)$ $\forall i$. Thus, $h_i(1)\equiv\psi$ $\forall i$ and $DS(t^*)=1-\psi-E(t^*)/E(0)$. 
Concerning the dynamics of shock propagation, we can easily study two limiting cases.
\begin{itemize}
 \item When $\psi\simeq0$, we can safely assume that no bank has failed after the first propagation step: $h_i(2)<1\;\forall i$. 
 We thus have $h_i(2)=\psi\{1+\sum_j[\lambda\Lambda_{ij}+\rho\gamma(1)\Upsilon_{ij}]\}$ $\forall i$ and, after some algebra, 
 $DS^*_{no-fail}=\psi C[\lambda+\rho\gamma(1)]/E(0)$, where $\gamma(1)=\rho\psi/(1-\rho\psi)$. 
 For $\tau\to0$, the dynamics stops at this stage as banks have already spread their distress once. For $\tau>0$, instead, the dynamics goes on and eventually banks start to default.
 \item When $\psi\simeq1$, all banks will ultimately fail after some rounds of distress propagation. In general, if at $t=t^*$ each bank has failed, 
 we get $h_i(t^*)=1$ $\forall i$ and $E(t^*)=0$, so that $DS^*_{full-fail}=1-\psi$. 
\end{itemize}

\begin{figure}[h!]
    \centering
    \includegraphics[width=\linewidth]{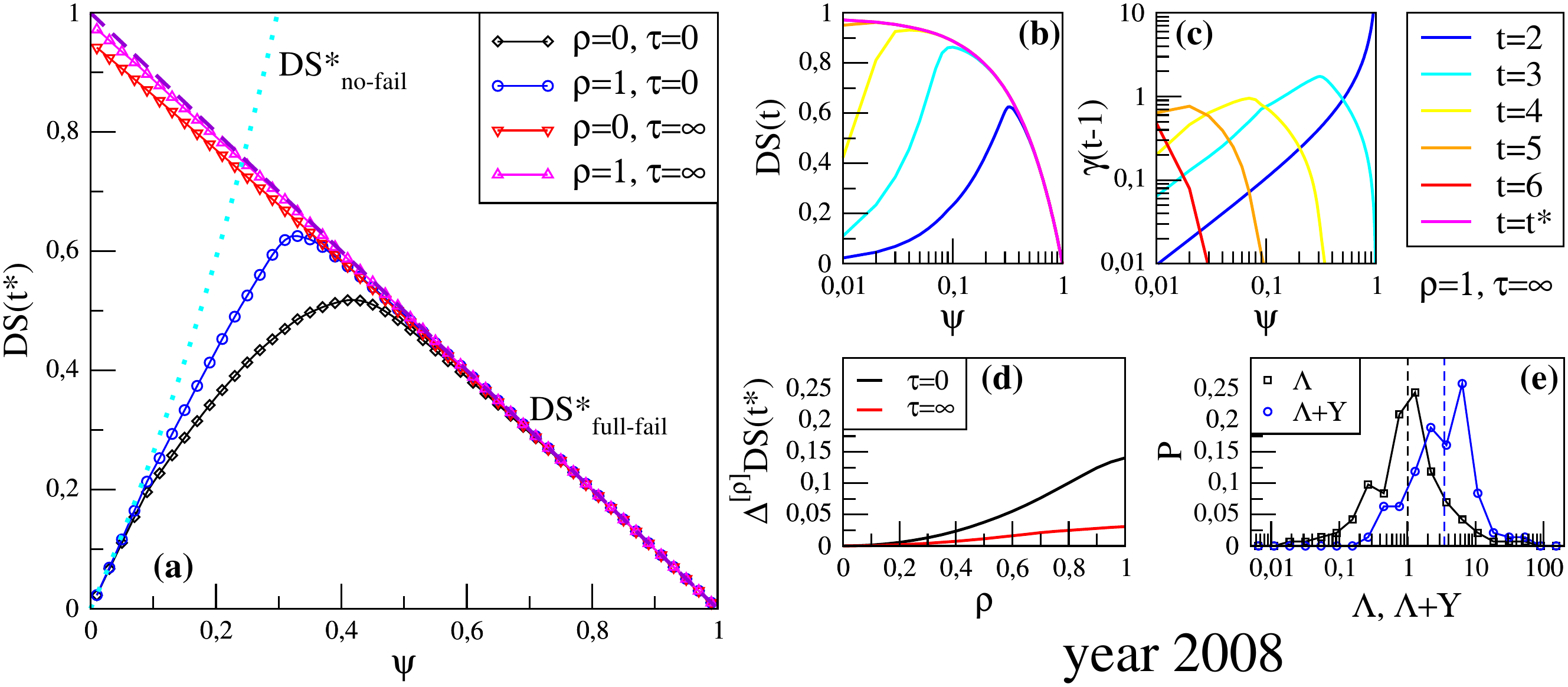}

    \vspace{0.5cm}

    \includegraphics[width=\linewidth]{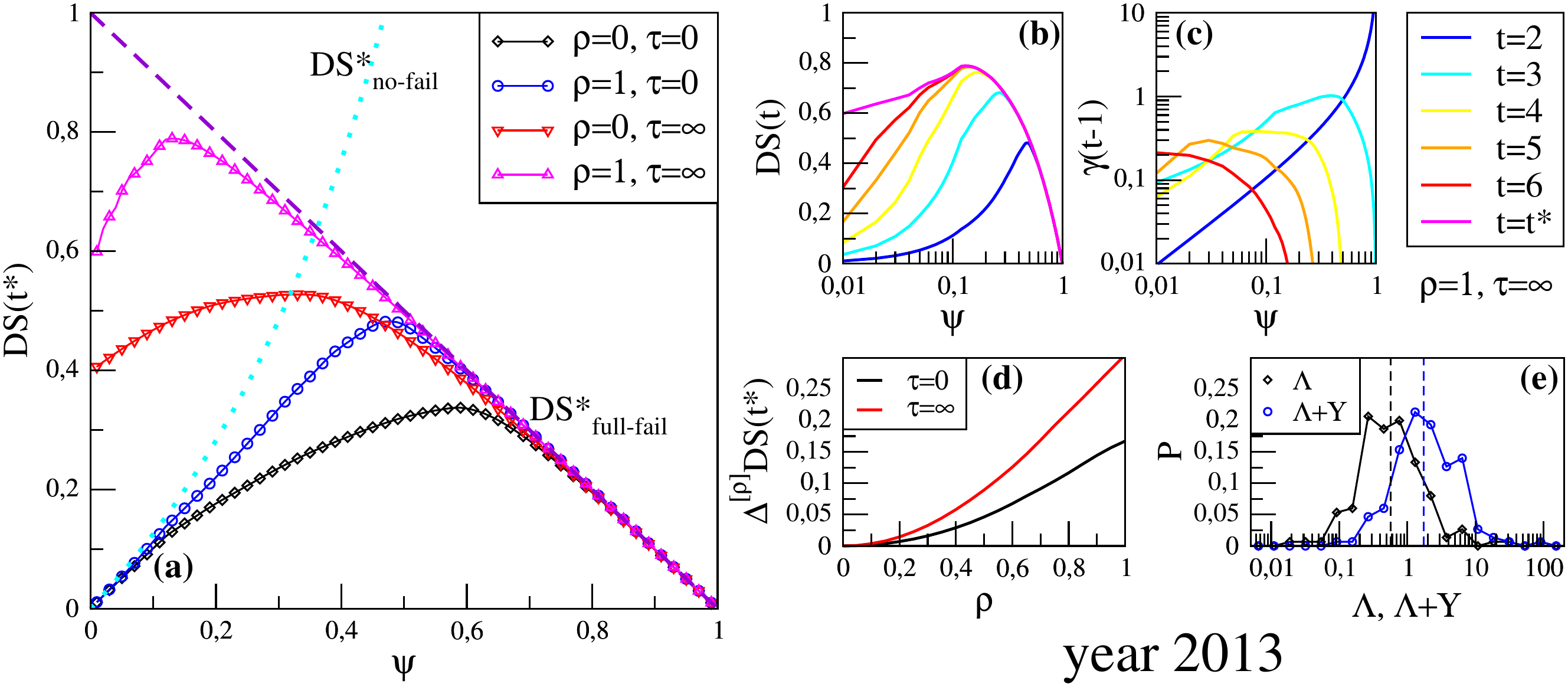}
\caption{Detailed results of group $DS$ Rank for years 2008 (upper panels) and 2013 (lower panels). 
(a) Stationary value $DS(t^*)$ as a function of the initial shock $\psi$, in the cases of maximal or vanishing strength of liquidity shocks ($\rho=1$ or $\rho=0$, respectively) 
and of instantaneous or vanishing damping of shocks ($\tau=0$ and $\tau=\infty$, respectively). (b,c) Temporal dynamics of the system: $DS$ and fire sale devaluation factor $\gamma$ 
as a function of $\psi$ for various iteration steps $t$, in the setting with liquidity shocks $\rho=1$ and multiple shocks propagation $\tau=\infty$. 
(d) Progressive effect of liquidity shocks: difference $\Delta^{[\rho]}DS(t^*)$ between $DS^{[\rho]}(t^*)$ (when banks must sell illiquid assets for a fraction $\rho$ of the lost funding) 
and $DS^{[0]}(t^*)$ (when no liquidity shocks occur), for $\psi$ maximizing the overall equity loss. 
(e) Probability distributions of banks leverages $\Lambda$ (black diamonds) and $\Lambda+\Upsilon$ (blue circles), with relative median given by a dashed vertical line.}
    \label{fig:group_0813}
\end{figure}

Figure \ref{fig:group_0813} shows the results of the group $DS$ dynamics for two temporal snapshots of our dataset: 2008 (the year of the global financial crisis) 
and 2013 (the most recent year at our disposal). Panels (a) summarize the stationary profile of $DS$ as a function of the initial shock $\psi$. 
Let us first consider the case $\tau=0$ of banks spreading their distress only once. As expected, for $\psi\simeq0$ the system is close to the configuration of $DS^*_{no-fail}$ with no bank defaults. 
Then, as $\psi$ increases, the overall equity loss grows, and after reaching its maximum value $DS(t^*)$ converges to the $DS^*_{full-fail}$ configuration with all bank defaults. 
The clear effect of liquidity shocks is that of increasing equity losses (especially in correspondence of the peak), and making the dynamics converge to $DS^*_{full-fail}$ for smaller initial shocks. 
Equity losses are however much heavier when subsequent rounds of shock propagation are allowed. In the extreme case $\tau\to\infty$, strikingly for year 2008 the dynamics converge to $DS^*_{full-fail}$ 
even if the initial distress is very small. By contrast, for year 2013 some banks survive, indicating a higher resilience of the system to financial distress. 
In any case, the dynamics needs more rounds to converge for small values of $\psi$, as shown in panels (b), and the fire sale devaluation factor $\gamma$ eventually assumes values close to 1, 
as shown in panels (c): large fire sales are triggered at least once during the evolution of the system. 
By denoting the stationary value of $DS$ Rank for a given value of $\rho$ as $DS^{[\rho]}(t^*)$, we see from panels (d) that the progressive effect of liquidity shocks, 
given by the difference $\Delta^{[\rho]}DS(t^*)=DS^{[\rho]}(t^*)-DS^{[0]}(t^*)$ is almost linear in $\rho$, and can reach values up to 0.3 for $\rho=1$. 
Finally, in order to gain insights on the reason why the resilience of the interbank market changed so significantly between the two considered years, 
we study the interbank leverage matrix, whose spectral properties determine the stability of the shock propagation dynamics~\cite{Bardoscia2015}. 
By definition, the interbank leverage ratio of a bank $i$ is given by $\Lambda_i=\sum_j\Lambda_{ij}$. Such value determines the bank resilience to credit shocks in the interbank market~\cite{Bardoscia2015}, 
with $\Lambda_i<1$ representing the ``safety'' condition for which that bank has enough capital to cover the losses due to the simultaneous defaults of all its debtor banks. 
Here, in order to account also for liquidity shocks, we introduced an {\em extended} interbank leverage ratio $\lambda\Lambda_i+\rho\Upsilon_i$, 
where $\Upsilon_i=\sum_j\Upsilon_{ij}$ and $\gamma=1$ according to the analysis of panels (c). The shape of the probability distributions of original and extended leverage ratios 
in the banking system, shown in panels (e), reveals how these values were much higher in 2008 than in 2013, resulting in a more fragile system in the previous case. 
Indeed, since the breakdown of the interbank market at the global financial crisis, some banks started to adopt a precautionary hoarding policy (due to concerns about future liquidity needs), 
and other banks in turn responded to the liquidity hoarding of others by adopting the same behavior \cite{Acharya2011,Berrospide2012}.
While precautionary motives still affect the interbank market nowadays (its total volume being steadily decreasing since 2007, see Tab. \ref{tab:stats}), 
the resulting reduced leverage ratios resulted in a more stable system.

\begin{figure}[h!]
    \centering
    \begin{minipage}{.3\textwidth}
        \centering
        \includegraphics[width=\linewidth]{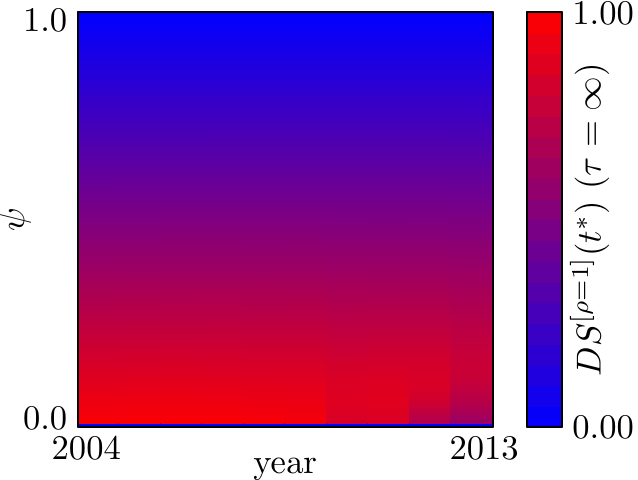}
    \end{minipage}
    \begin{minipage}{0.3\textwidth}
        \centering
        \includegraphics[width=\linewidth]{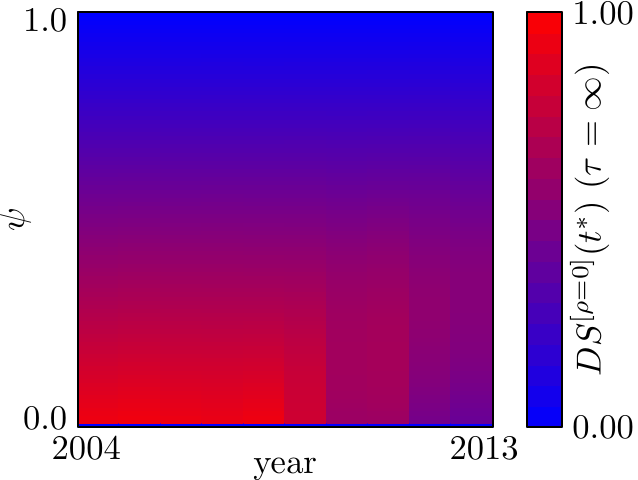}
    \end{minipage}
    \begin{minipage}{0.3\textwidth}
        \centering
        \includegraphics[width=\linewidth]{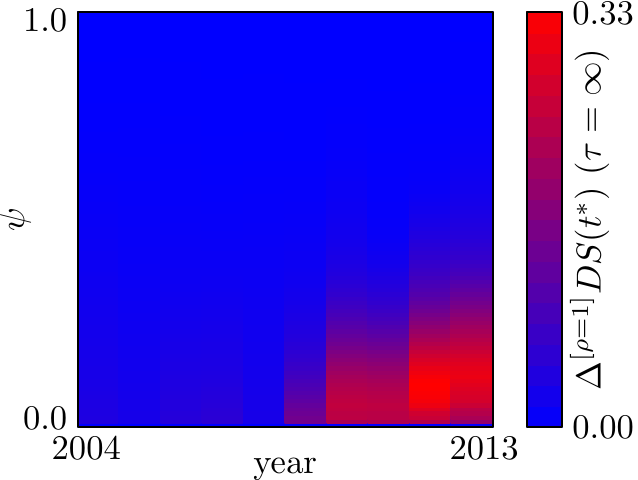}
    \end{minipage}
    \begin{minipage}{.3\textwidth}
        \centering
        \includegraphics[width=\linewidth]{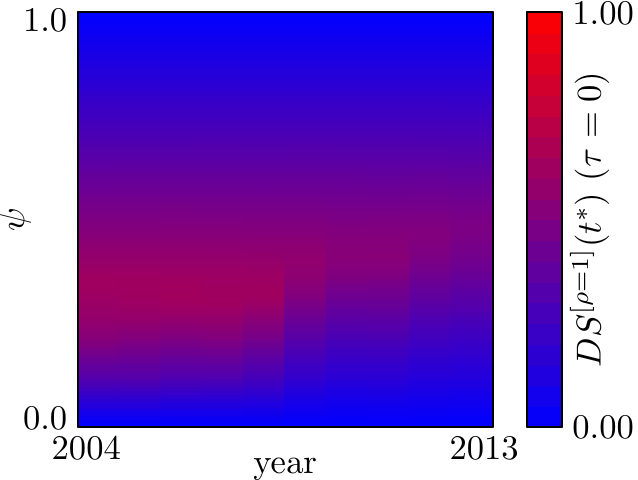}
    \end{minipage}
    \begin{minipage}{0.3\textwidth}
        \centering
        \includegraphics[width=\linewidth]{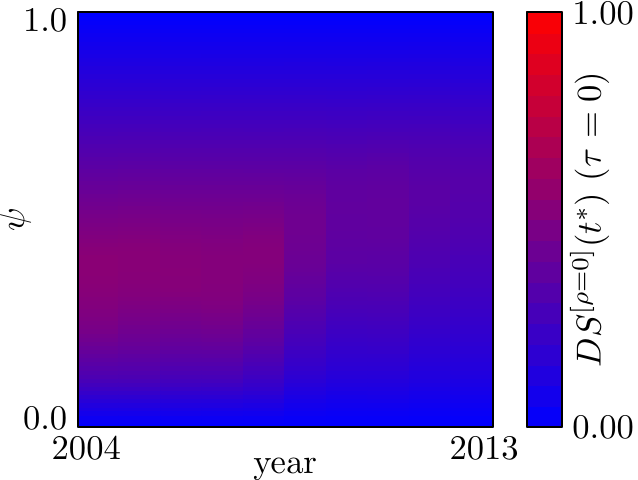}
    \end{minipage}
    \begin{minipage}{0.3\textwidth}
        \centering
        \includegraphics[width=\linewidth]{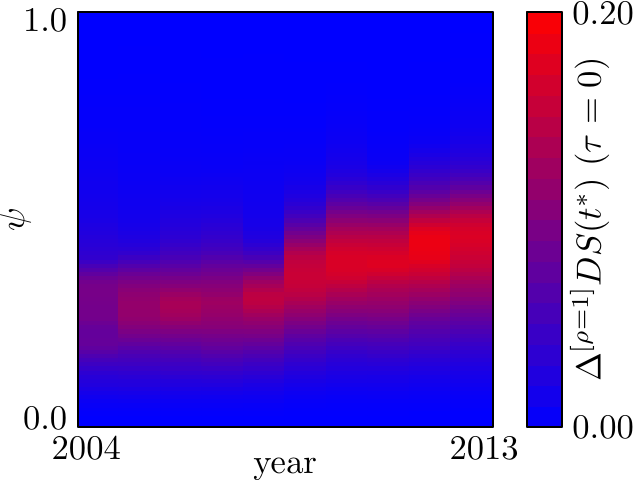}
    \end{minipage}
\caption{Heat maps of group $DS(t^*)$ as a function of $\psi$ and of the year. Upper plots refer to the case of no damping ($\tau=\infty$) and lower plots to the case of instantaneous damping ($\tau=0$). 
Then, from left to right, plots show $DS(t^*)$ with liquidity shocks and $\rho=1$, no liquidity shocks $\rho=0$, and the difference between these values.} 
    \label{fig:group_HM}
\end{figure}

While detailed results for other years can be found in the supplementary material, we can grasp the general systemic risk properties of the interbank market 
over the time span considered (2004-2013) from Figure \ref{fig:group_HM}, which shows color maps of the stationary group $DS$ values for the various years and as a function of the initial shock $\psi$. 
For $\tau=0$, $DS(t^*)$ keeps its ``inverse U'' shape for all years. Yet, while in the absence of liquidity shocks systemic risk substantially decreases after 2008, 
these shocks do lead to an increase in equity losses that shows up especially in late years. A similar behavior is observed even more clearly for $\tau\to\infty$, 
as in this case the distance from the full-fail scenario after 2008 is dramatically reduced by the presence of liquidity shocks. 
Overall, these results point to the importance of considering liquidity shocks when assessing potential equity losses in the financial system. 
Notably, after 2008 liquidity risk has become more prominent for higher values of the initial shock: in complex network terminology, 
the system has become more {\em robust} for low $\psi$, yet more {\em fragile} for high $\psi$.

\subsection*{Individual DS Rank}

We now move to the case in which the initial shock to the market corresponds to the failure of an individual bank $u$: 
$E_u(1)=0\Rightarrow h_u(1)=1$ and $E_i(1)=E_i(0)\Rightarrow h_i(1)=0$ $\forall i\neq u$. To make this initial condition explicit, 
we use the notation $X(t|u)$ for the value of a generic quantity $X$ at step $t$ an when $u$ is the initially shocked bank. 
We thus have $DS(t^*|u)=1-E(t^*|u)/E(0)-\nu_u$. Using this framework, we can compute two bank-specific risk indicators:
\begin{itemize}
 \item {\em Impact} $I$, given by the relative equity loss experienced by the market after the initial default of that bank:
$$I_u=1-\frac{E(t^*|u)}{E(1)}=\frac{E(0)-E_u(0)-E(t^*|u)}{E(0)-E_u(0)}=\frac{DS(t^*|u)}{1-\nu_u}.$$
 \item {\em Vulnerability} $V$, namely the relative equity loss for that bank averaged over the initial defaults of all other banks:
$$V_u=\left\langle\frac{E_u(0)-E_u(t^*|j)}{E_u(0)}\right\rangle_{j\neq u}=\left\langle h_u(t^*|j)\right\rangle_{j\neq u}.$$
\end{itemize}
 
\begin{figure}[h!]
    \centering
    \includegraphics[width=0.75\linewidth]{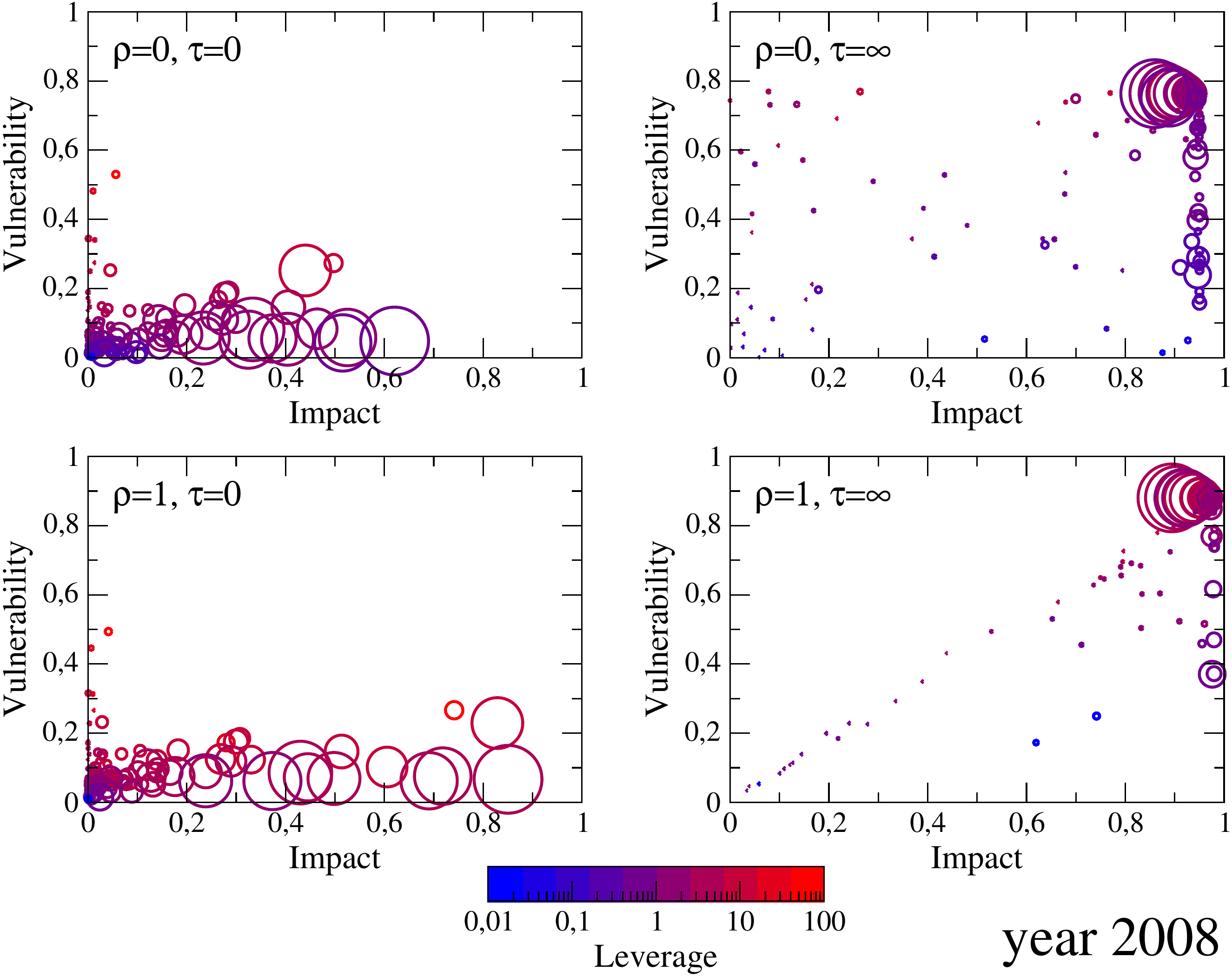}
    
    \vspace{0.5cm}
    
    \includegraphics[width=0.75\linewidth]{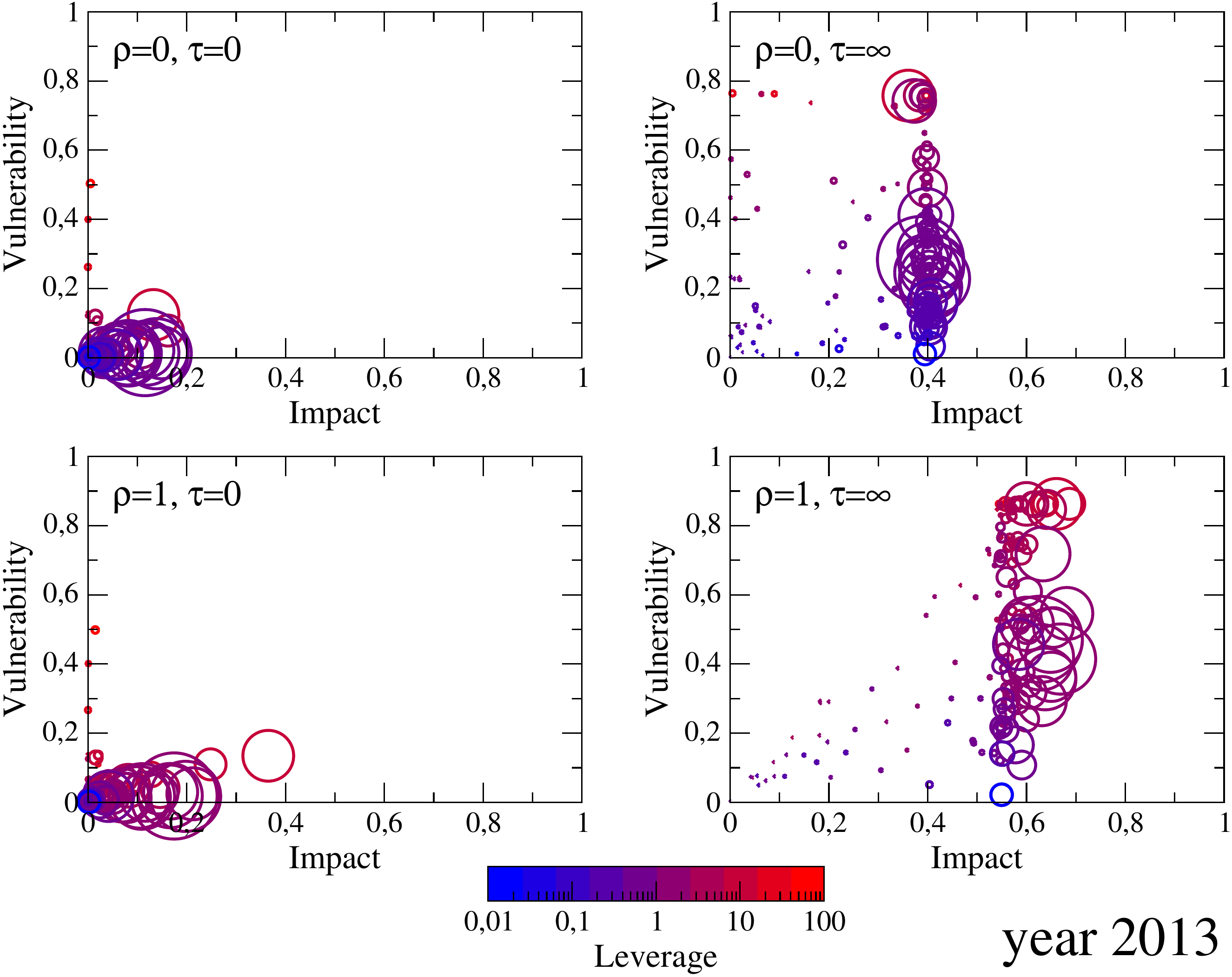}
\caption{Scatter plot of impact and vulnerability values for each bank in years 2008 (upper panels) and 2013 (lower panels), 
in the cases of maximal or vanishing strength of liquidity shocks ($\rho=1$ or $\rho=0$, respectively) 
and of instantaneous or vanishing damping of shocks ($\tau=0$ and $\tau=\infty$, respectively). 
The size of the bubble is non-linearly proportional to the initial equity of the corresponding bank: $\mbox{size}_i=10^{-2}\sqrt{E_i(0)/2}$. 
The color of the contour is instead given by the value of the generalized leverage $\lambda\Lambda_i+\rho\Upsilon_i$.}
    \label{fig:single_0813}
\end{figure}

Figure \ref{fig:single_0813} shows scatter plots of impact and vulnerability of individual banks again for years 2008 and 2013 
(plots corresponding to other years can be found in the supplementary material). 
First note that, as expected, vulnerability clearly increases with the leverage ratio of the bank, whereas, impact is more related to the bank size (as measured by its initial equity). 
According to our data, interestingly the banks with the highest leverage have relatively small size, and despite being very vulnerable they lack the potential to affect the market significantly. 
Concerning the different shock propagation scenarios, in general we observe that the systemic impact of banks drops significantly from 2008 to 2013, 
while the decrease of systemic vulnerability is less evident but for multiple shock propagation $\tau\to\infty$. 
The effect of liquidity shocks is in turn that of increasing both measures. In particular, in year 2013 these shocks cause an increase of 100\% (for $\tau=0$) and of 50\% (for $\tau\to\infty$) 
for the impact of the largest banks, again underlying the importance of fire sale spillovers for a more robust assessment of systemic risk.

We conclude by remarking that, according to our analysis, impact indeed correlates with size but is not a monotonic function of it: some banks can be more impactful 
on the system than others with larger equity (or leverage). Thus, the characteristics of the interbank network are important to determine the extent of financial contagion: 
with respect to banking systems regulation, relying on balance sheet constraints may be insufficient, and should be supplemented by considerations for the structure of the network~\cite{Krause2012}.

\section*{Discussion}

In this work we have focused on the interbank lending market, the network of financial interlinkages resulting from overnight loans between banks. 
This network represent an important sector of the whole financial system in terms of traded volumes~\cite{Smaga2016}, and is crucial for banks to cope with liquidity fluctuations 
and meet reserve requirements~\cite{Allen2014}. In spite of its role, the interbank market is quite fragile, as intrasystem cash fluctuations alone have the potential 
to lead to systemic defaults~\cite{Smaga2016}, and exceptional external shocks can lead the market to a complete drought~\cite{Brunnermeier2009a}. 
In order to quantify potential equity losses in interbank markets resulting from exogenous shocks, building on Debt Rank~\cite{Battiston2012}, here we have defined a systemic risk metric 
that combines the dynamics of credit and liquidity shocks propagation~\cite{Lau2009}: the Debt-Solvency Rank. By applying this framework to a dataset of 183 European banks from 2004 to 2013, 
we have showed that fire sale spillovers do increase the overall equity loss by a factor up to 50\%, and almost double the individual systemic impact of banks---especially in years after 2008. 
All these evidences point to the importance of considering liquidity risk in network-based stress tests. 
We also document that the interbank market was extremely fragile also before 2008, as in those years even the smallest initial shock would have caused all banks to default 
(provided no taming of shocks propagation). By contrast, after the crisis the market became able to absorb an increasing amount of financial distress. 

In our framework, fire sales occur as banks sell their illiquid assets to recover lost fundings, 
and these sales happen {\em simultaneously} with the propagation of credit and funding shocks. In this respect, our approach is different from~\cite{Battiston2015X}, 
where fire sales occur only {\em after} the distress propagation process and because of the target leverage policy adopted by banks. 
Note also that in our model we consider assets sales solely by banks, but these spillovers are likely to trigger additional sales by other financial institutions, 
which would depress prices even more. In this respect, in principle the $DS$ Rank could be combined with the {\em aggregated vulnerability} framework of \cite{Greenwood2015,Duarte2015} 
in order to consider liquidity contagion through both direct exposures and common asset holdings. Additionally, we have not considered in our model any liquidity hoarding behavior for banks. 
Overall, the effect of funding shocks might be much more relevant than what can be assessed through the $DS$ Rank alone. Still, our approach can provide a useful lower bound 
for the potential impact of these shocks on the financial system, and indirectly on the real economy~\cite{Gabbi2015} 
(as banks suffering from liquidity shortages end up reducing their investments and affecting the economic growth). 
In any event, our analysis supports the thesis that liquidity requirements on financial institutions 
may be as effective as capital requirements in hindering contagious failures~\cite{Cifuentes2005,Nier2008}.

\section*{Acknowledgements}
We thank Marco Bardoscia, Paolo Barucca and Guido Caldarelli for useful discussions, and Marco D'Errico for sharing the data. 
G.C. acknowledges support from the EU projects GROWTHCOM (FP7-ICT, n. 611272), MULTIPLEX (FP7-ICT, n. 317532), SIMPOL (FP7-ICT, n. 610704) and DOLFINS (H2020-EU.1.2.2., n. 640772). 
The funders had no role in study design, data collection and analysis, decision to publish, or preparation of the manuscript.

%

\newpage
\section*{Supplementary Information --- Group DS Rank}

\begin{figure}[h!]
\begin{center}
\includegraphics[width=0.3\textwidth]{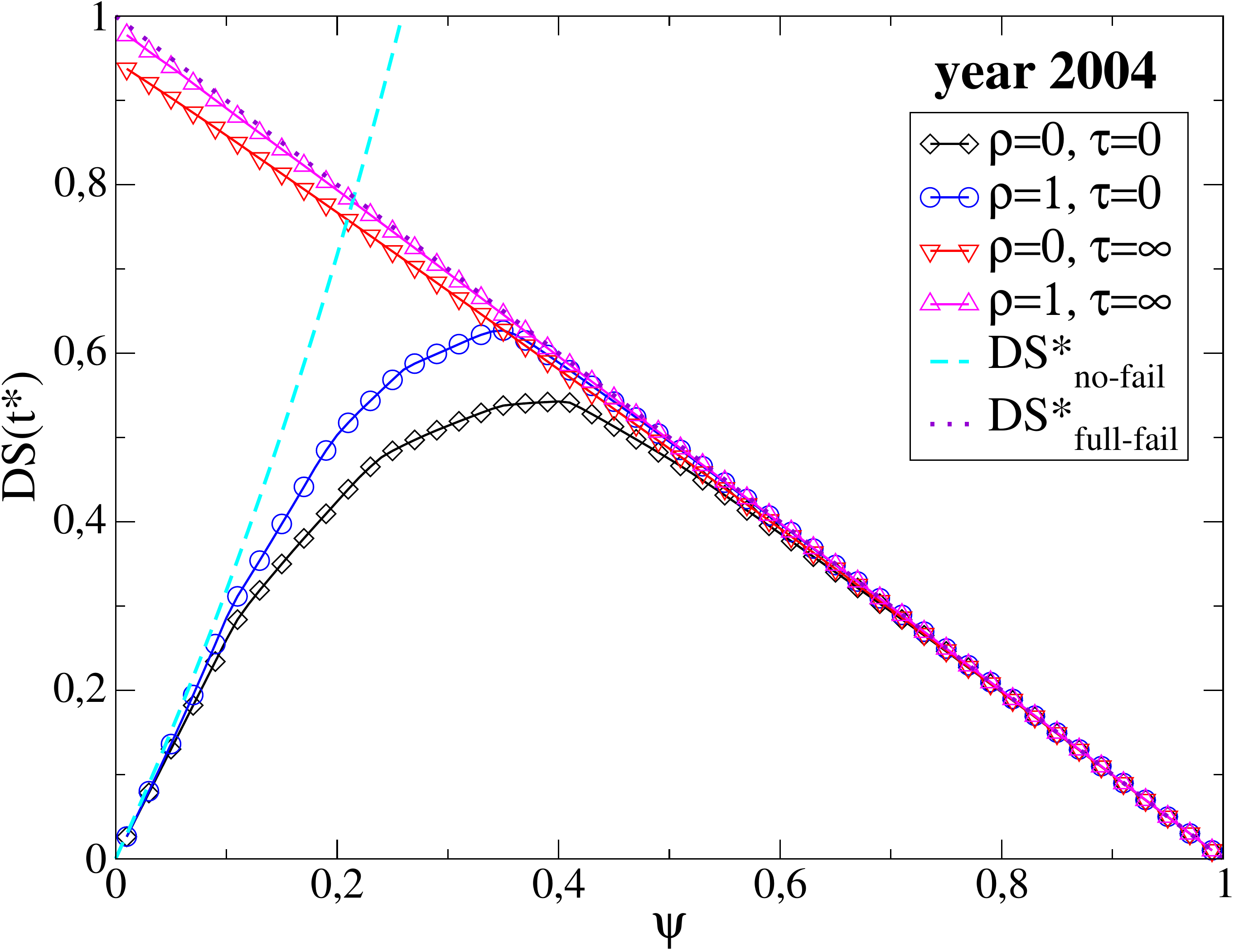}
\includegraphics[width=0.3\textwidth]{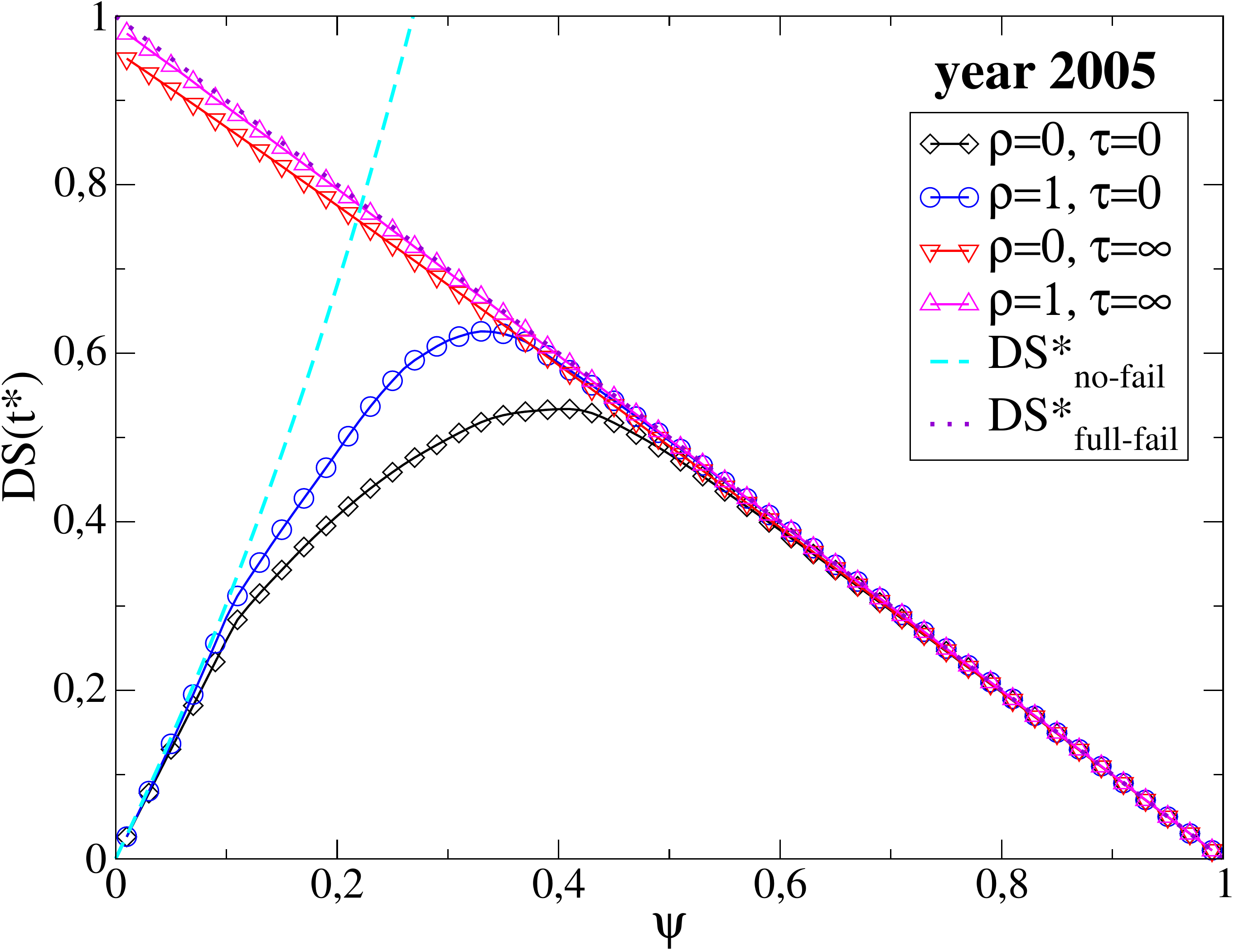}\\
\includegraphics[width=0.3\textwidth]{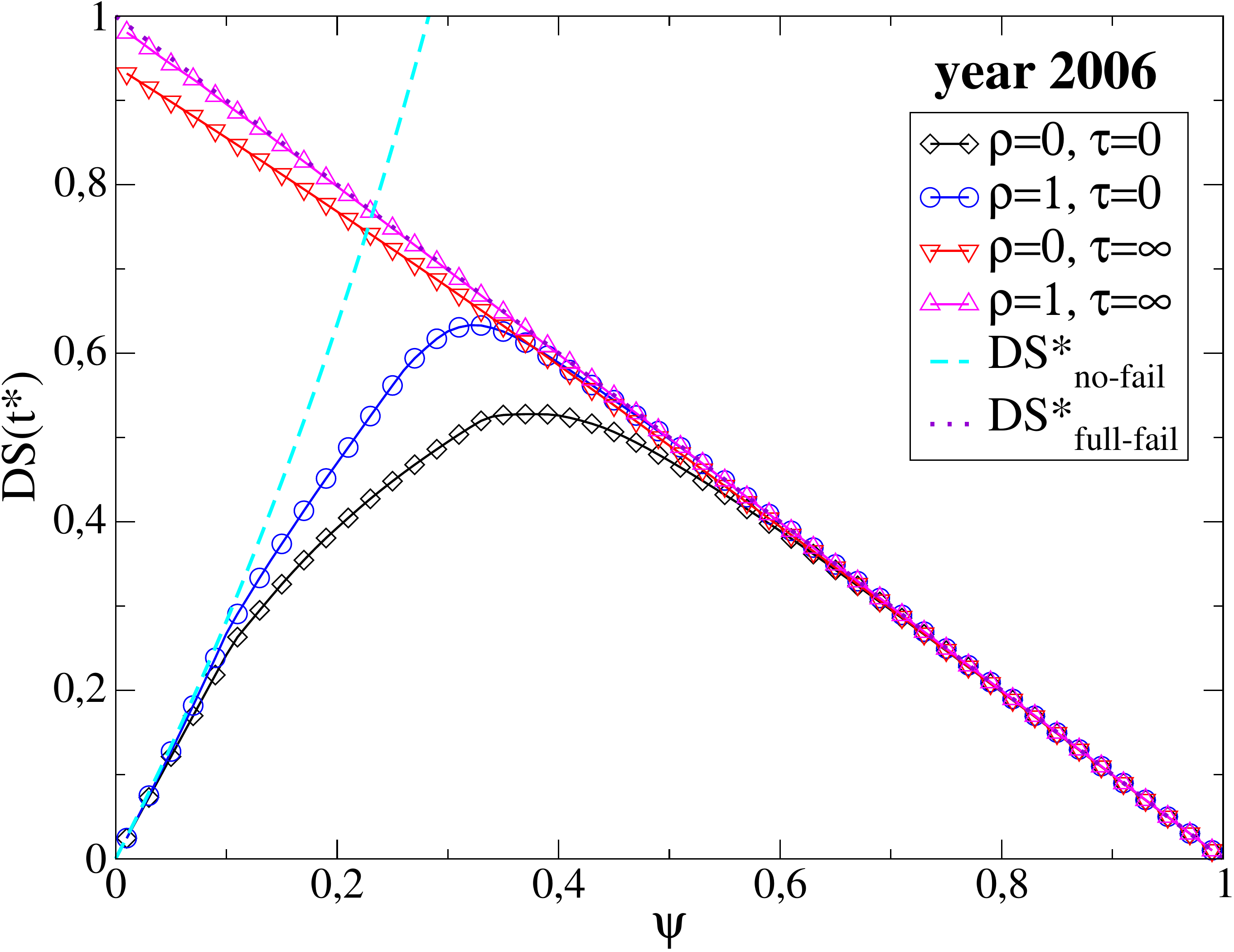}
\includegraphics[width=0.3\textwidth]{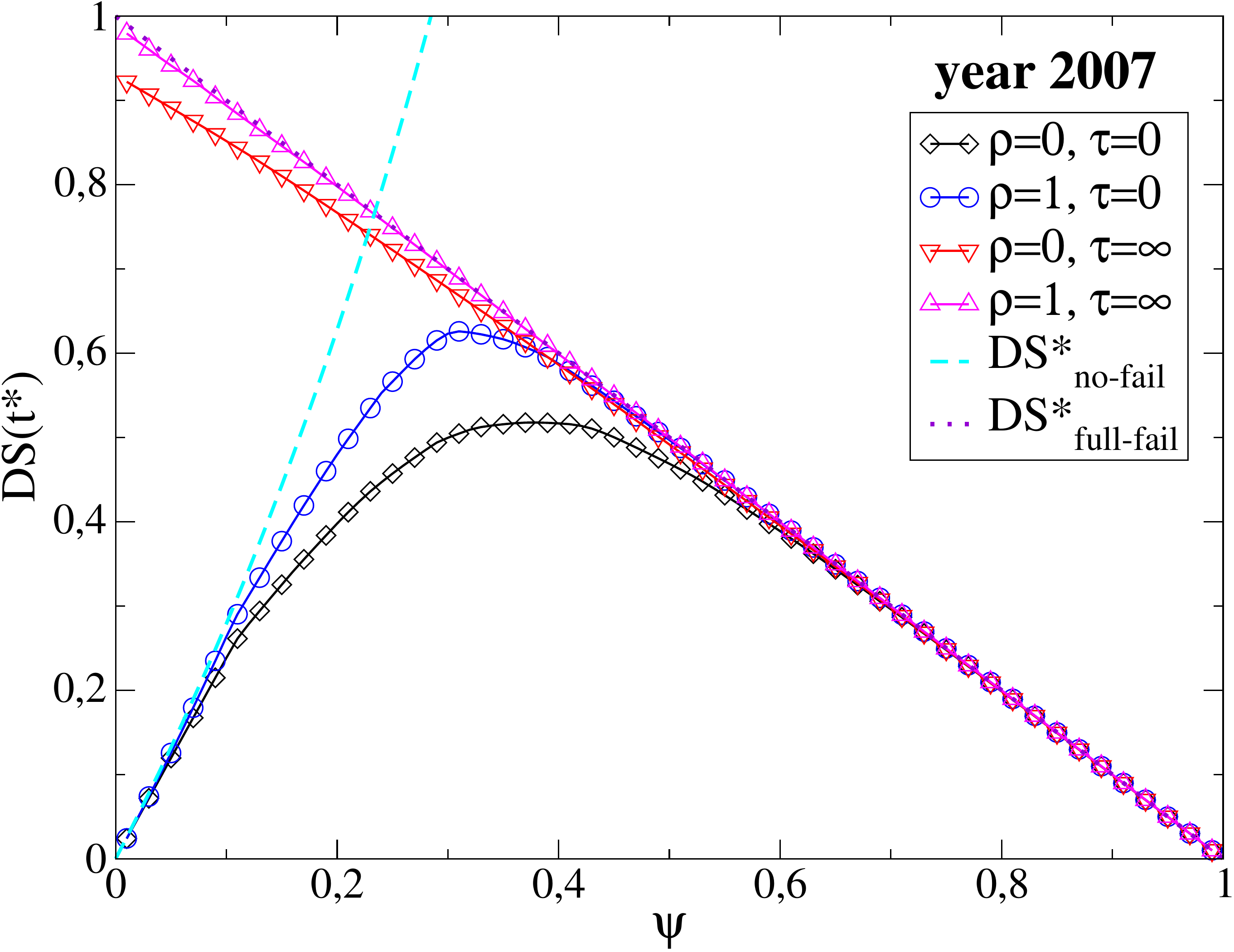}\\
\includegraphics[width=0.3\textwidth]{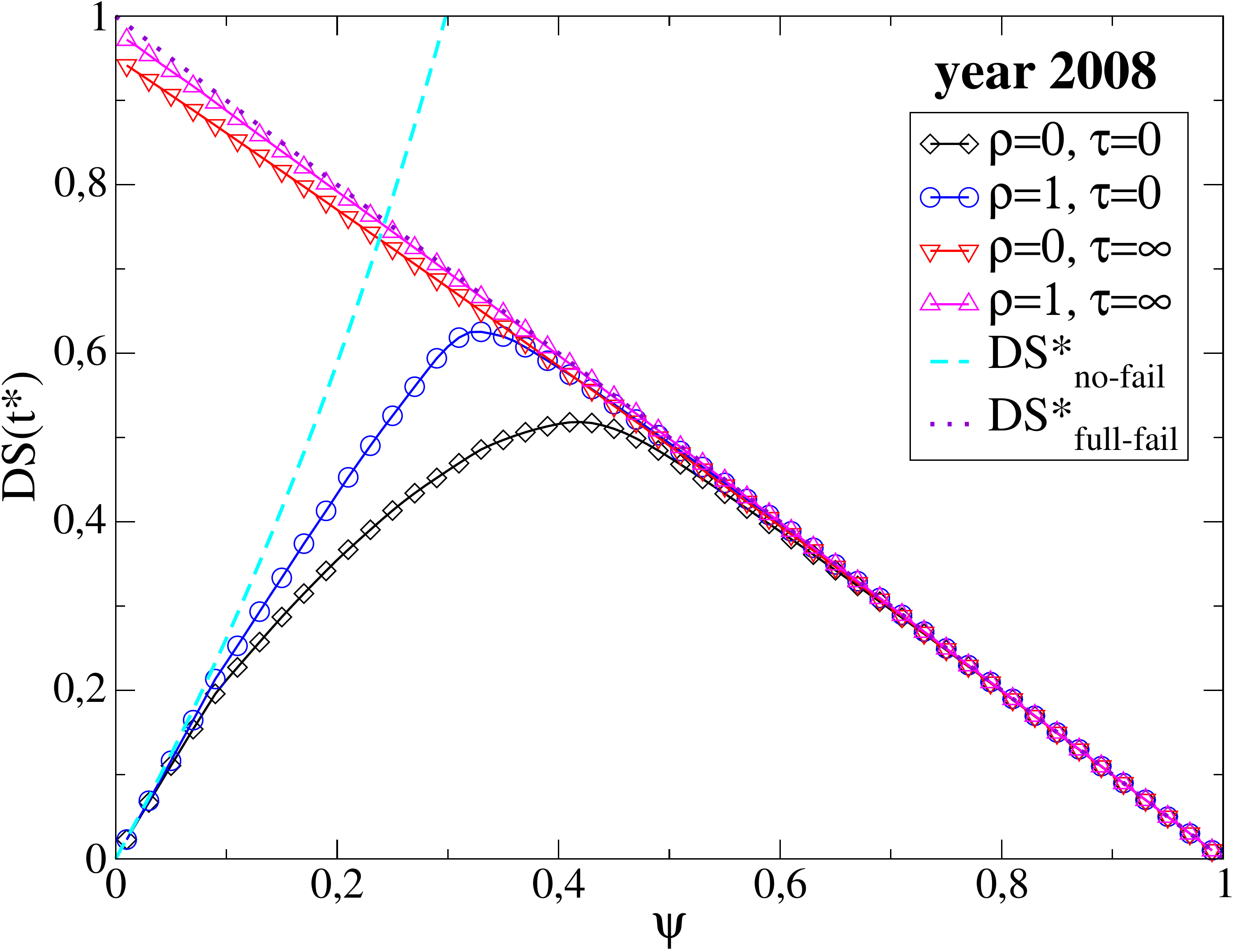}
\includegraphics[width=0.3\textwidth]{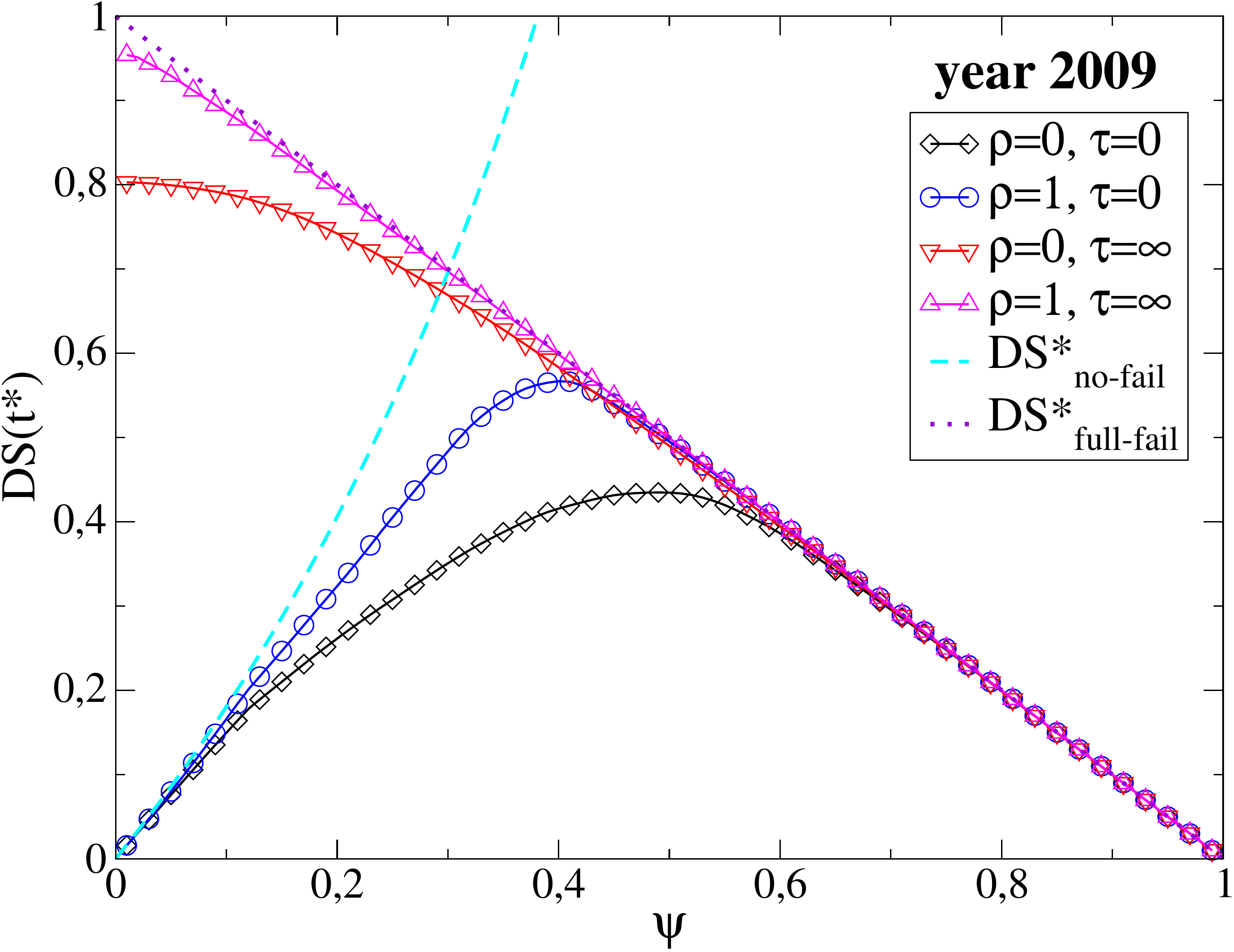}\\
\includegraphics[width=0.3\textwidth]{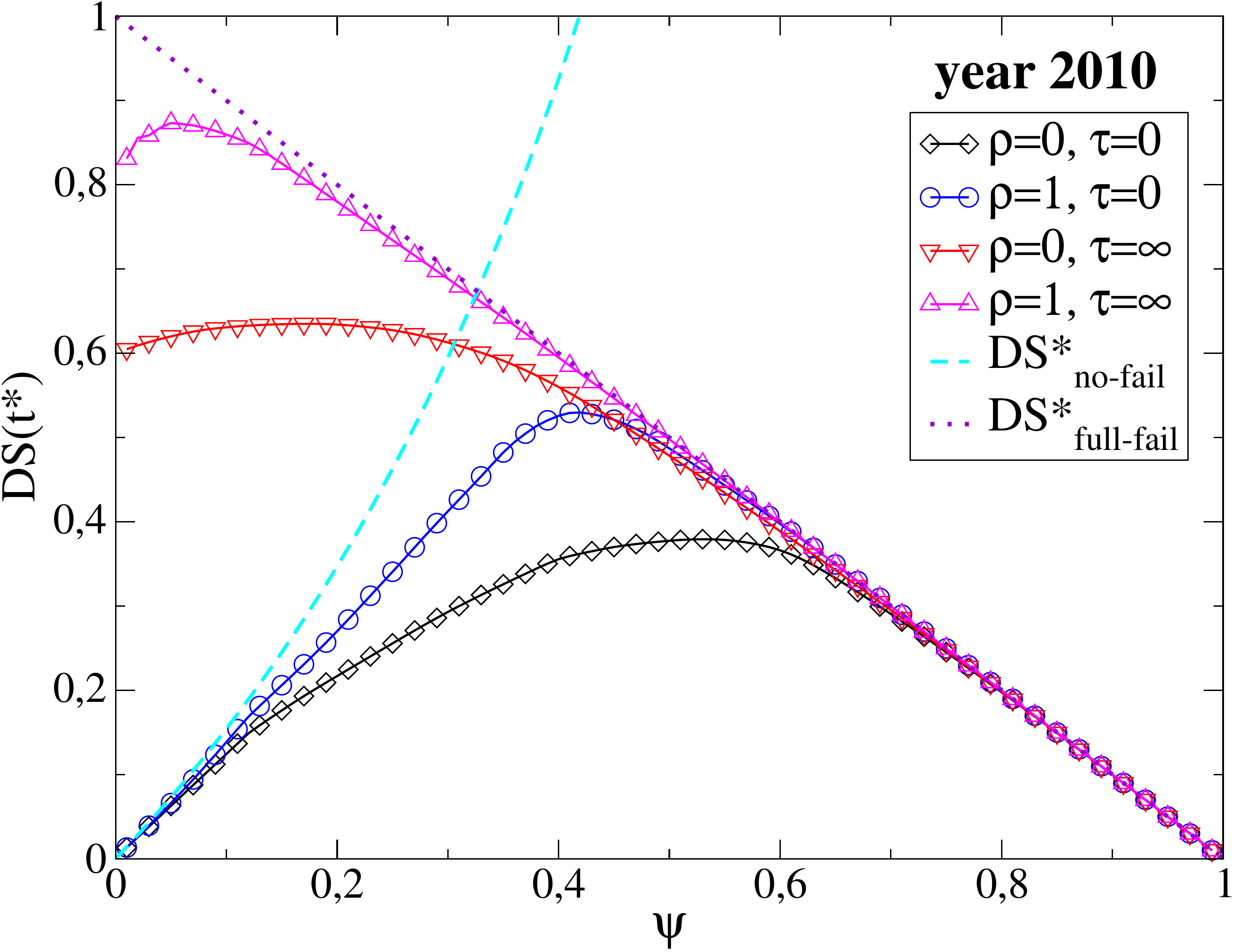}
\includegraphics[width=0.3\textwidth]{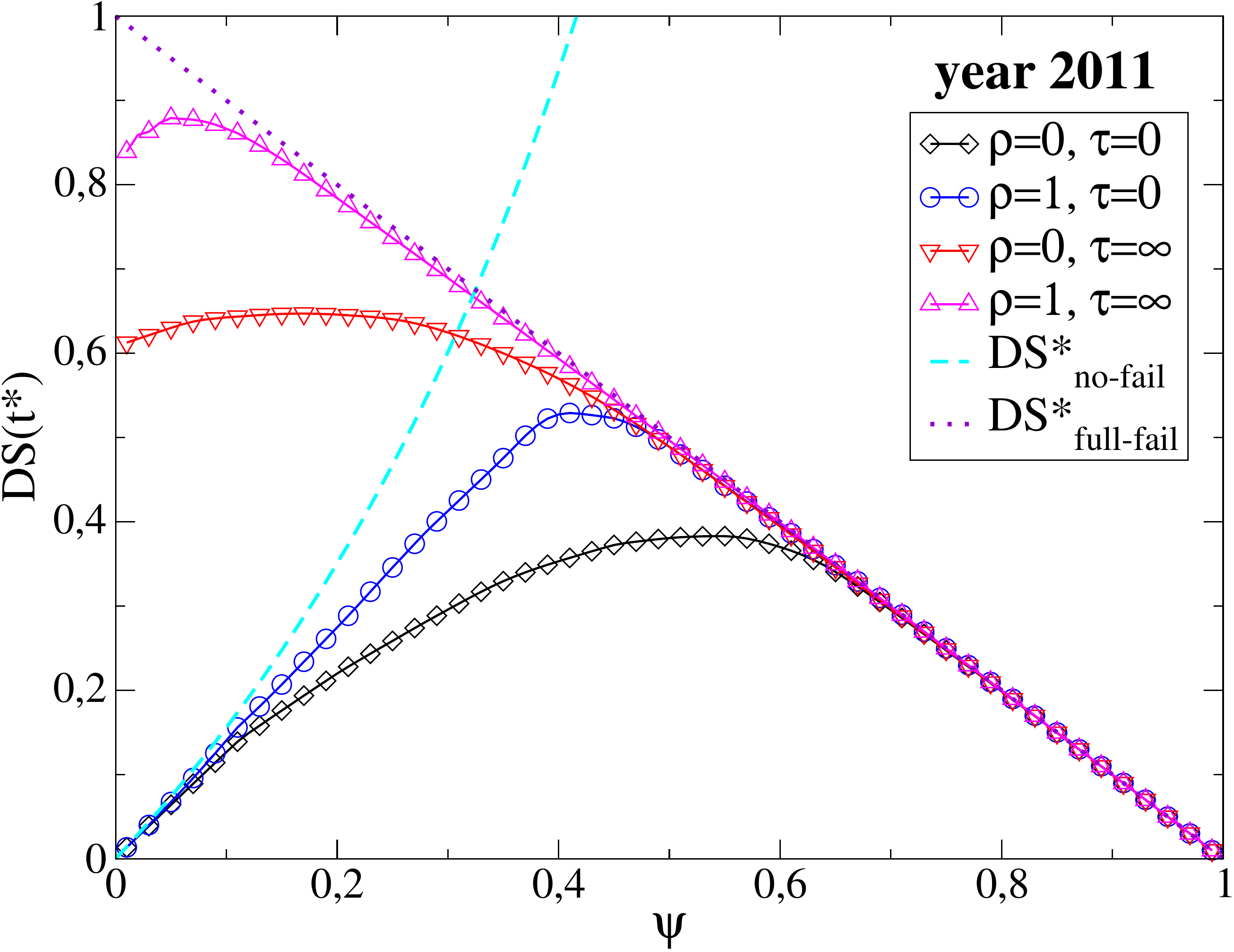}\\
\includegraphics[width=0.3\textwidth]{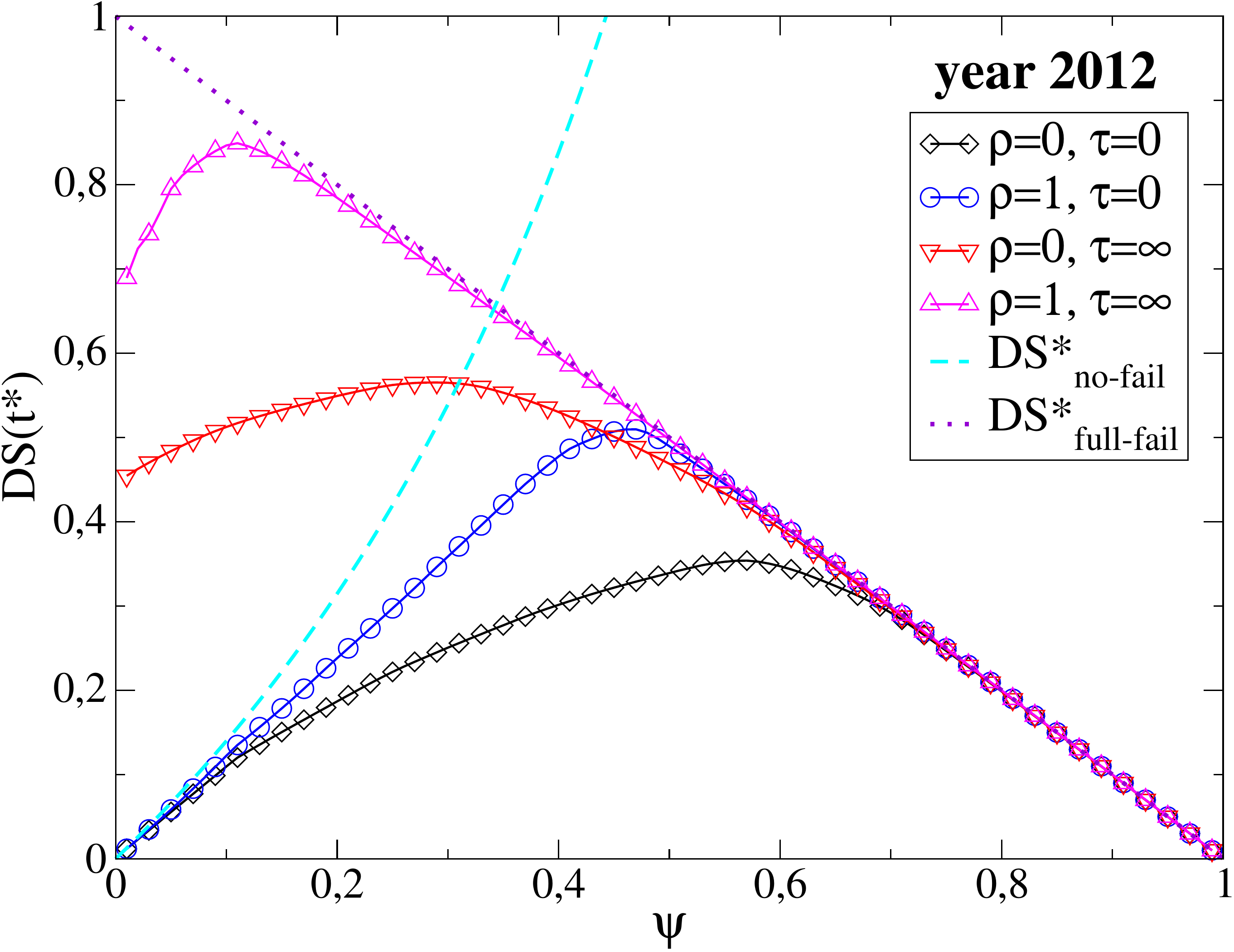}
\includegraphics[width=0.3\textwidth]{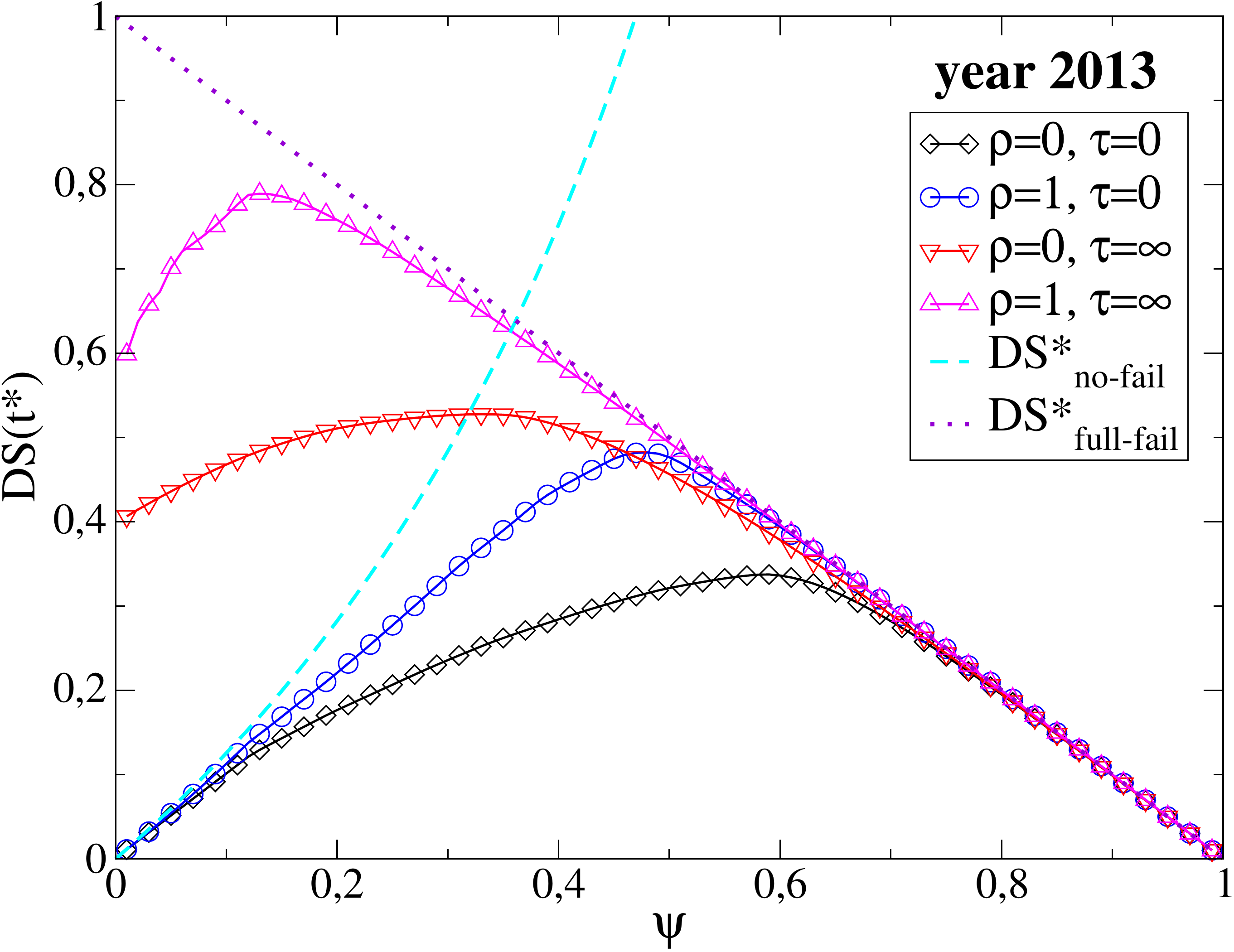}
\end{center}
\end{figure}

\newpage
\section*{Supplementary Information --- Individual DS Rank (part 1)}
\begin{figure}[h!]
\begin{center}
\includegraphics[width=0.45\textwidth]{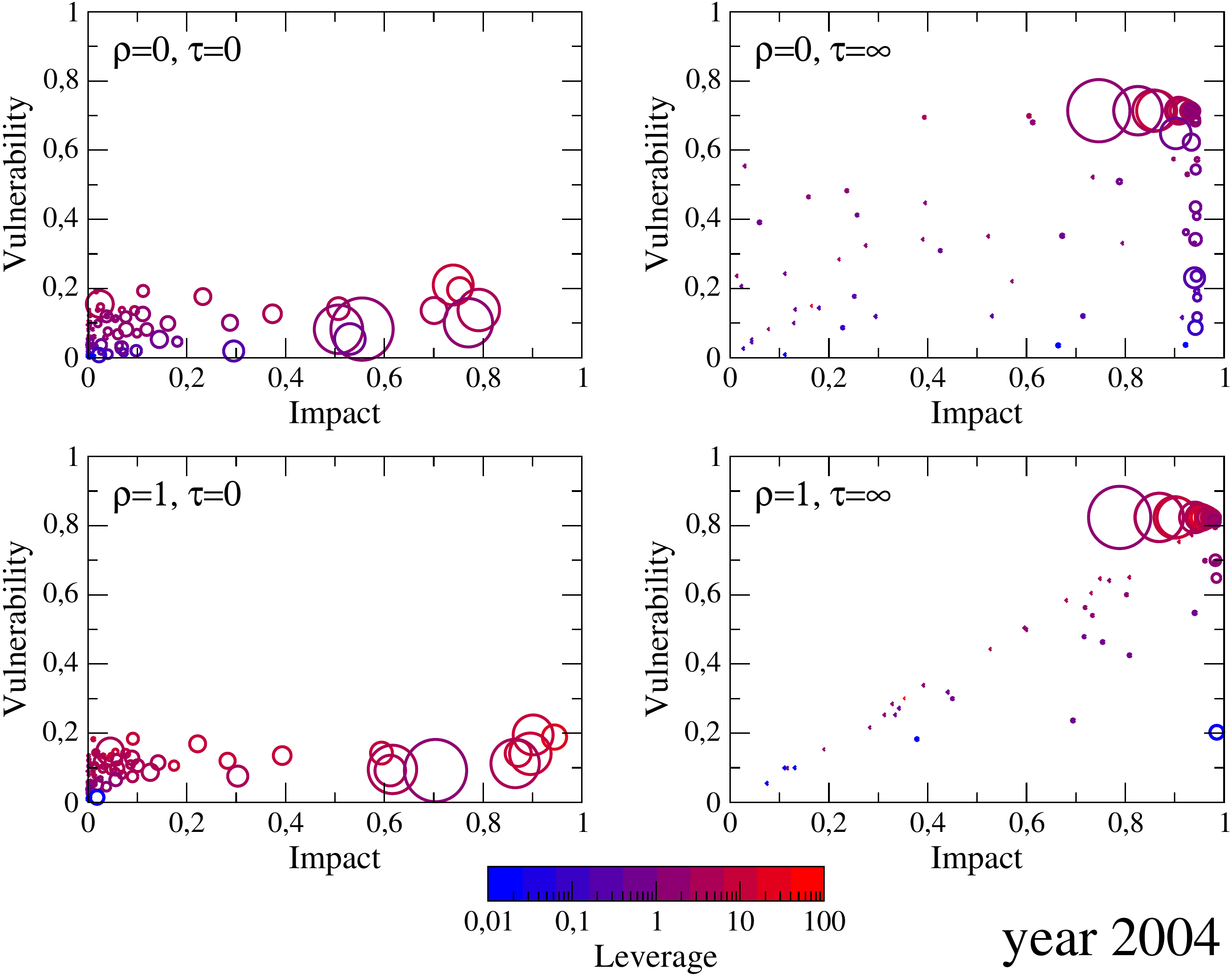}
\includegraphics[width=0.45\textwidth]{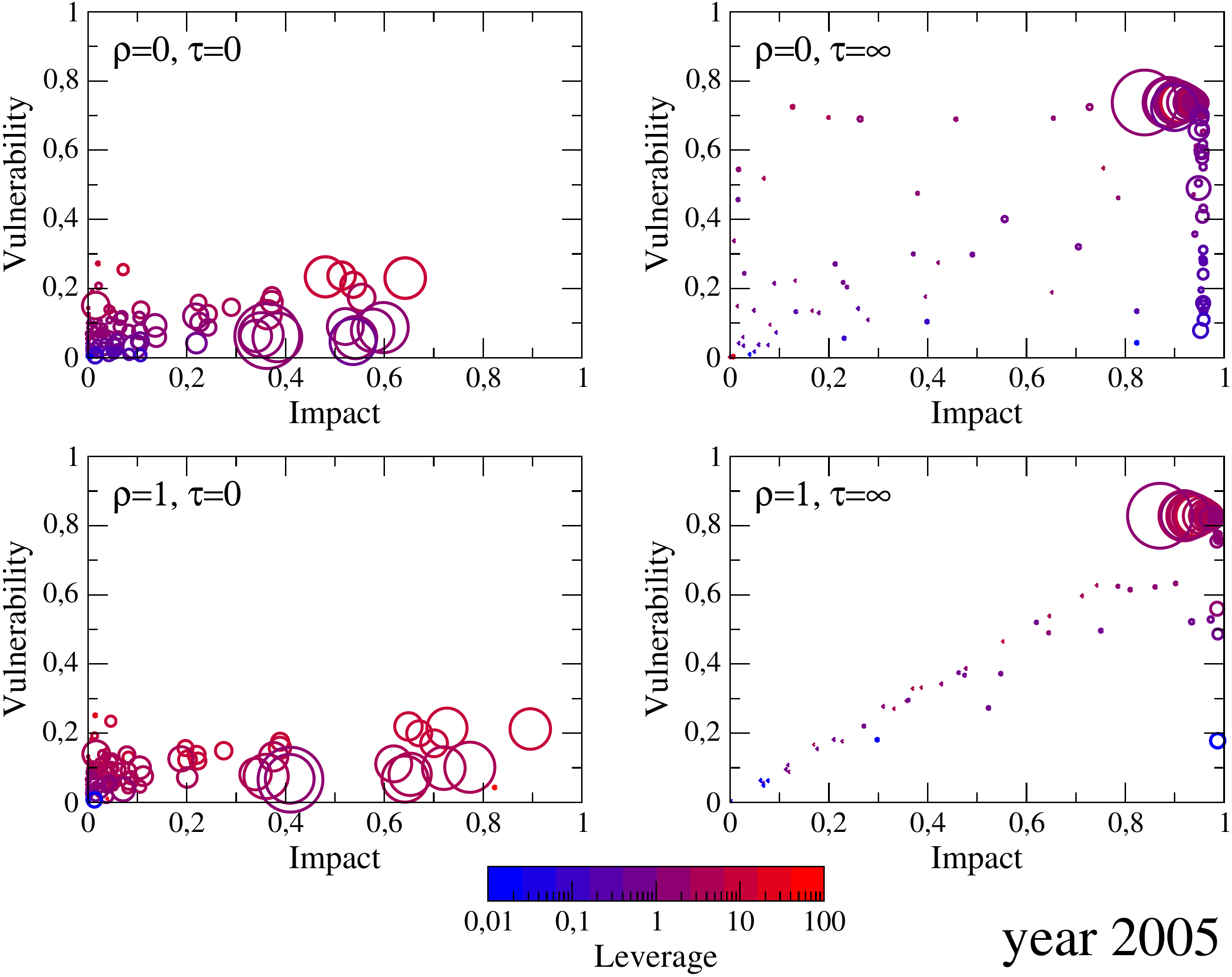}\\
\vspace{0.5cm}
\includegraphics[width=0.45\textwidth]{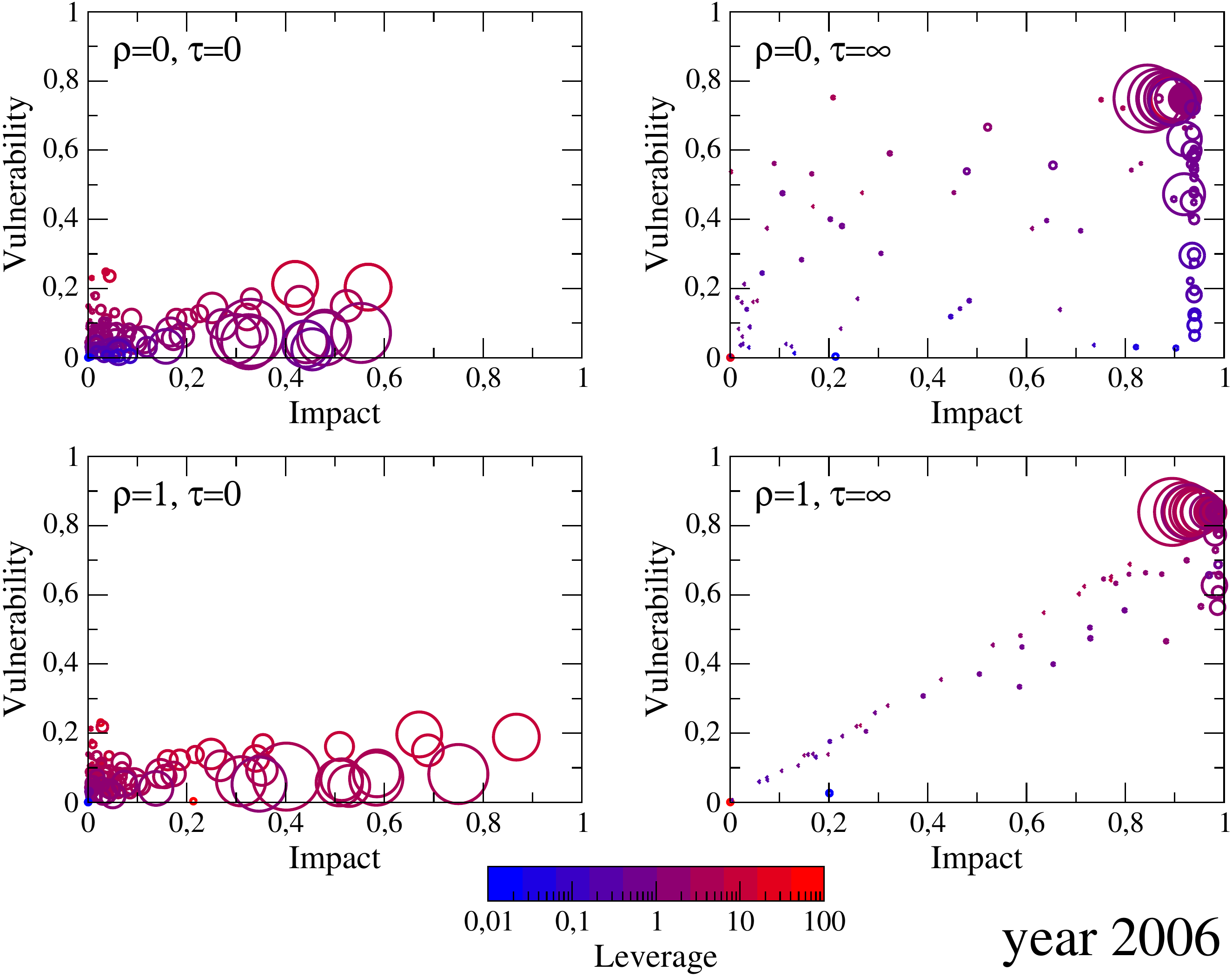}
\includegraphics[width=0.45\textwidth]{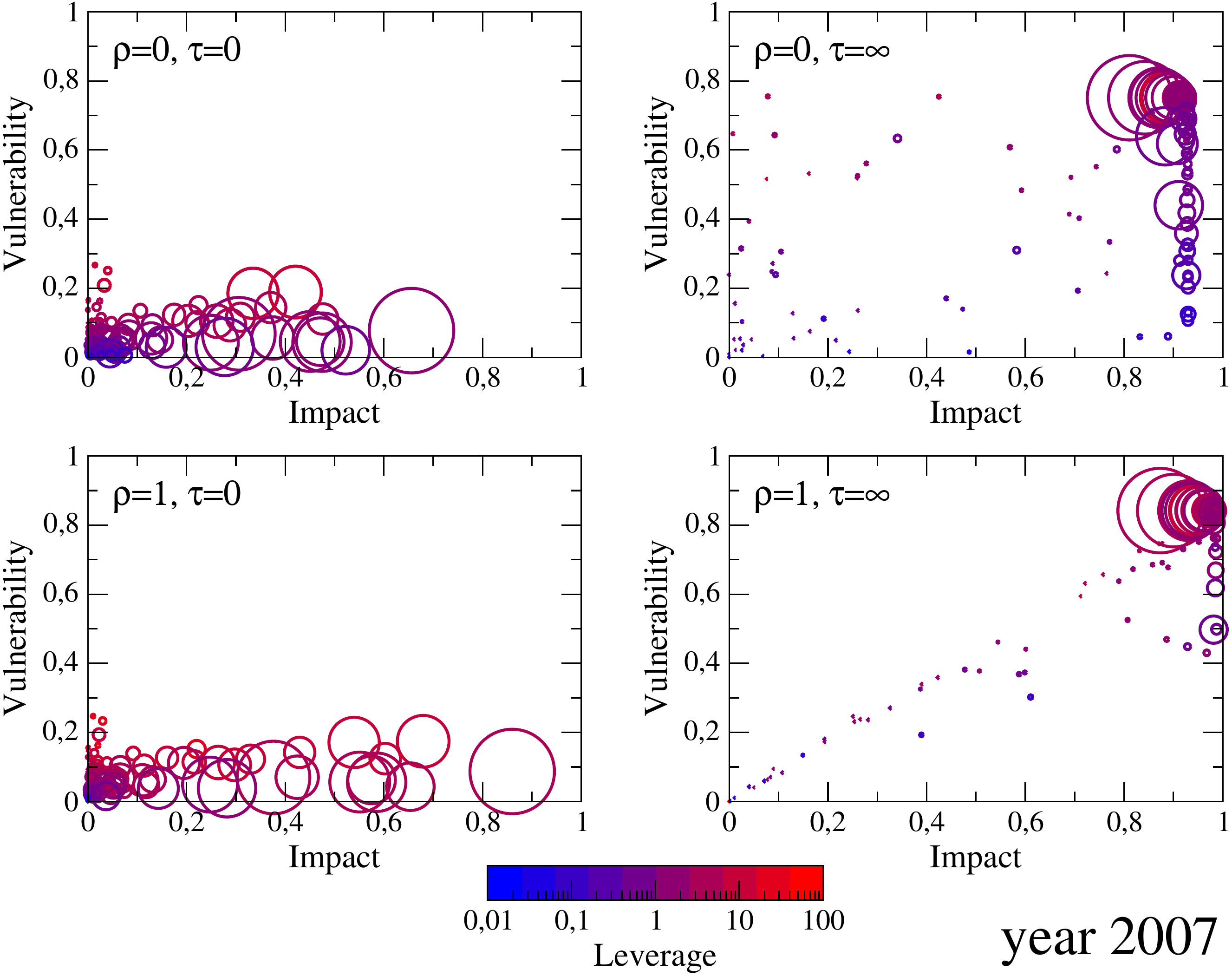}\\
\vspace{0.5cm}
\includegraphics[width=0.45\textwidth]{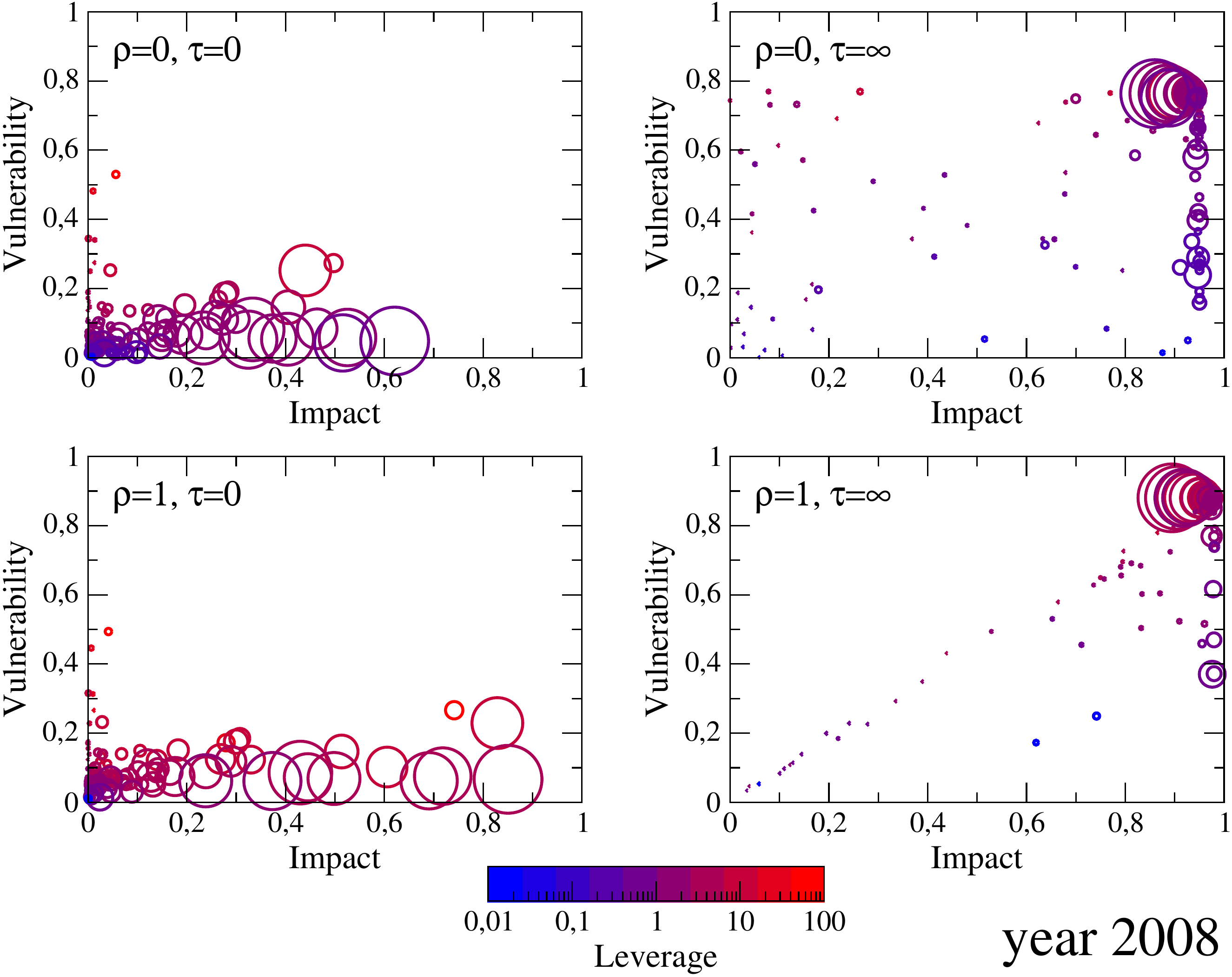}
\end{center}
\end{figure}

\newpage
\section*{Supplementary Information --- Individual DS Rank (part 2)}
\begin{figure}[h!]
\begin{center}
\includegraphics[width=0.45\textwidth]{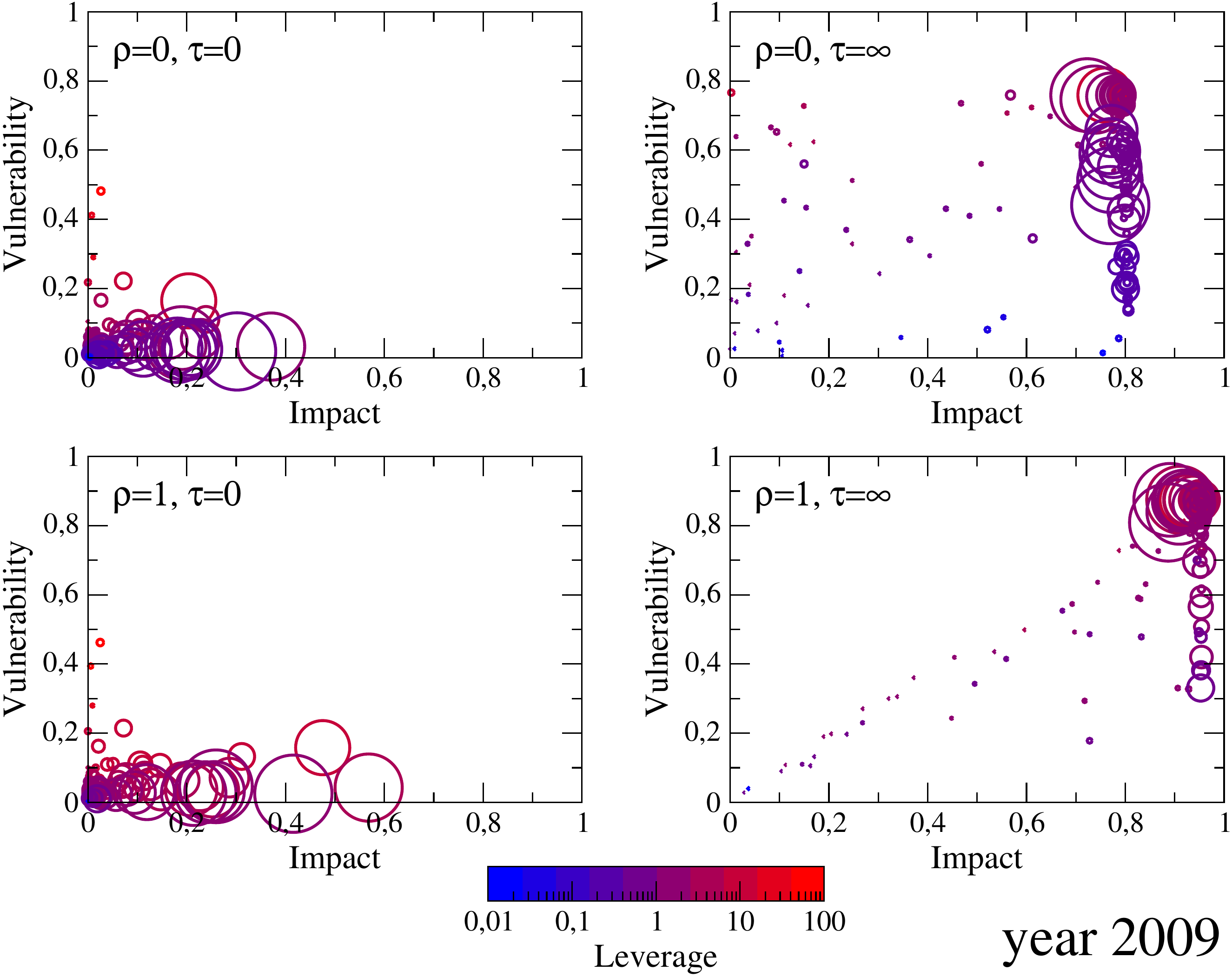}
\includegraphics[width=0.45\textwidth]{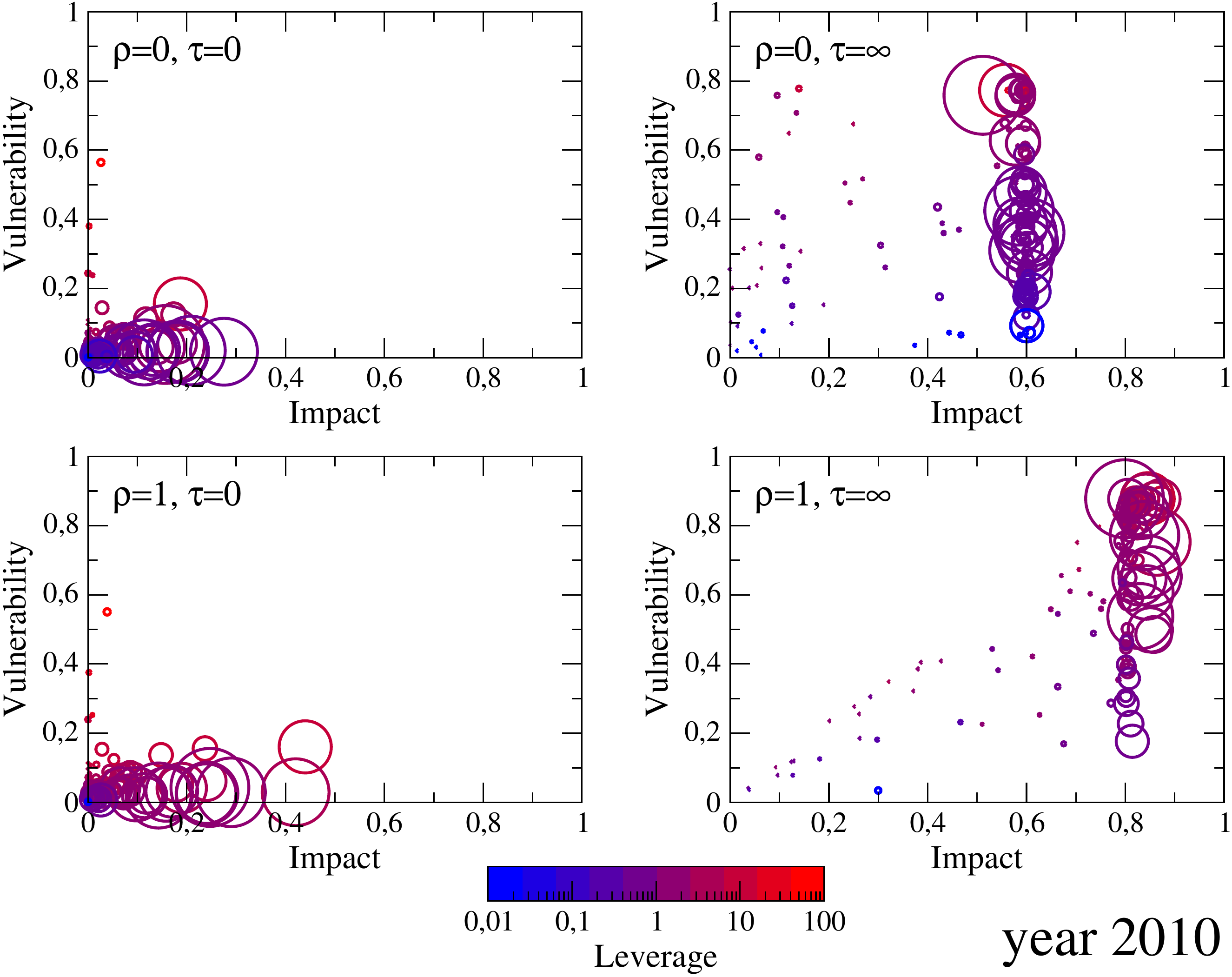}\\
\vspace{0.5cm}
\includegraphics[width=0.45\textwidth]{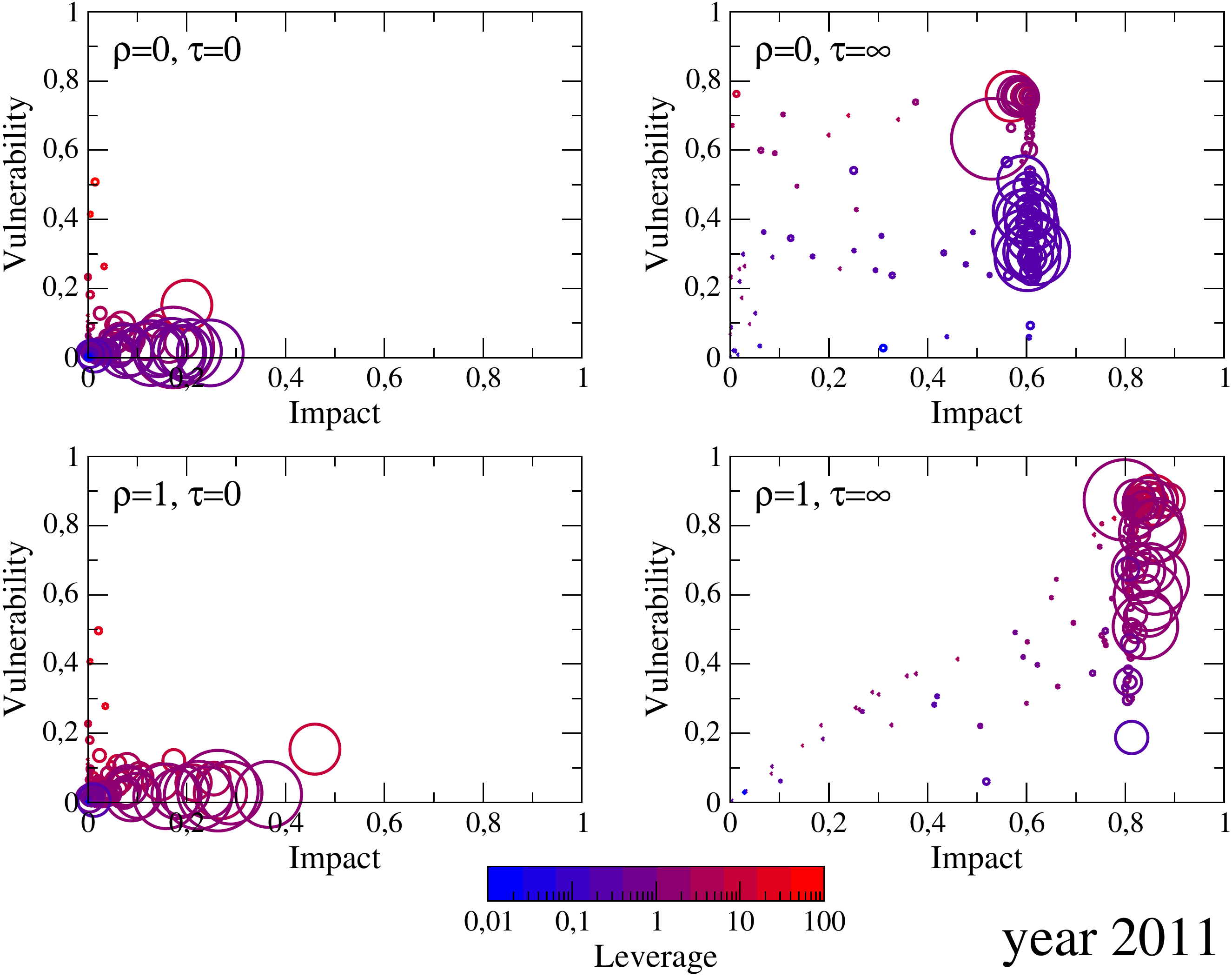}
\includegraphics[width=0.45\textwidth]{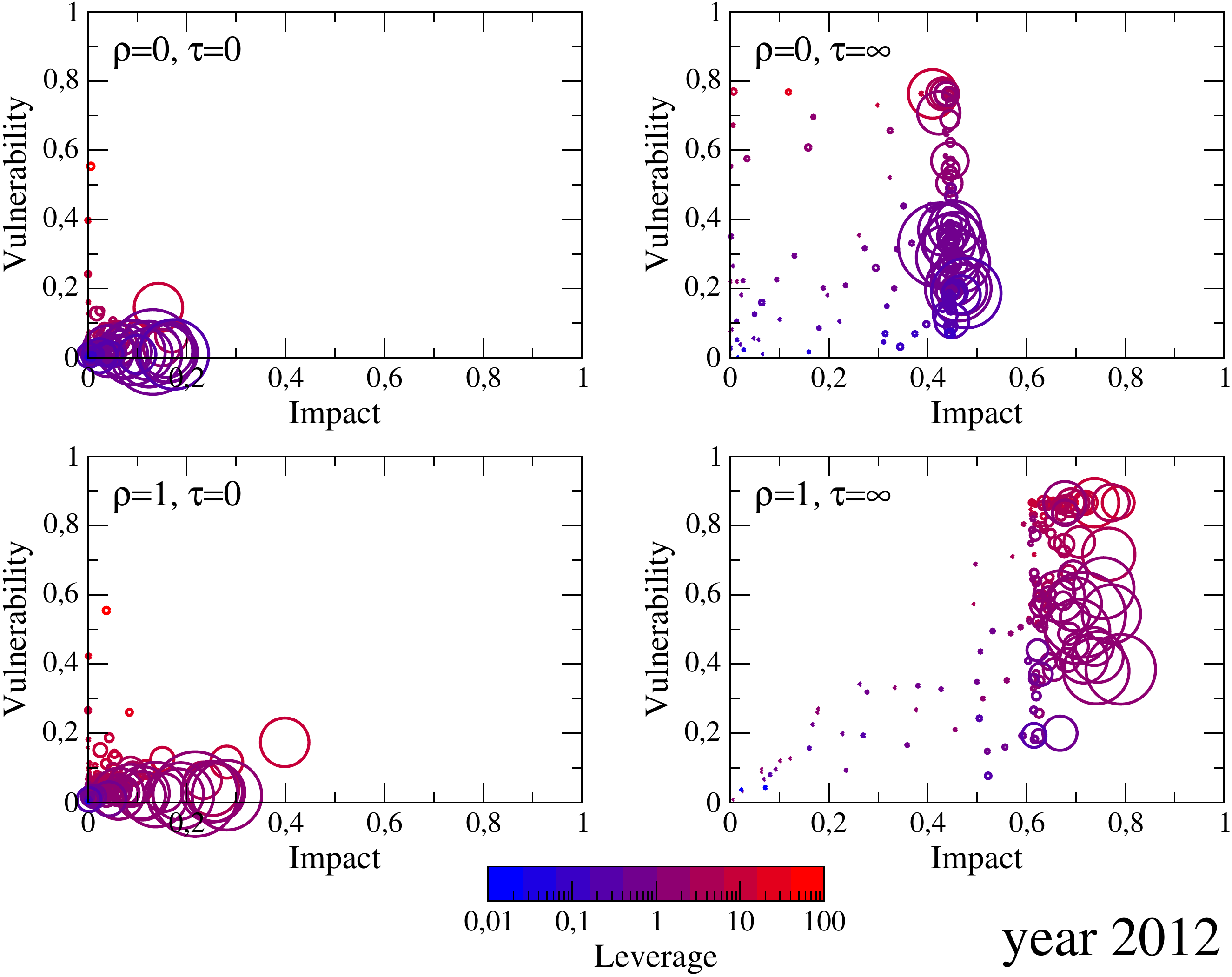}\\
\vspace{0.5cm}
\includegraphics[width=0.45\textwidth]{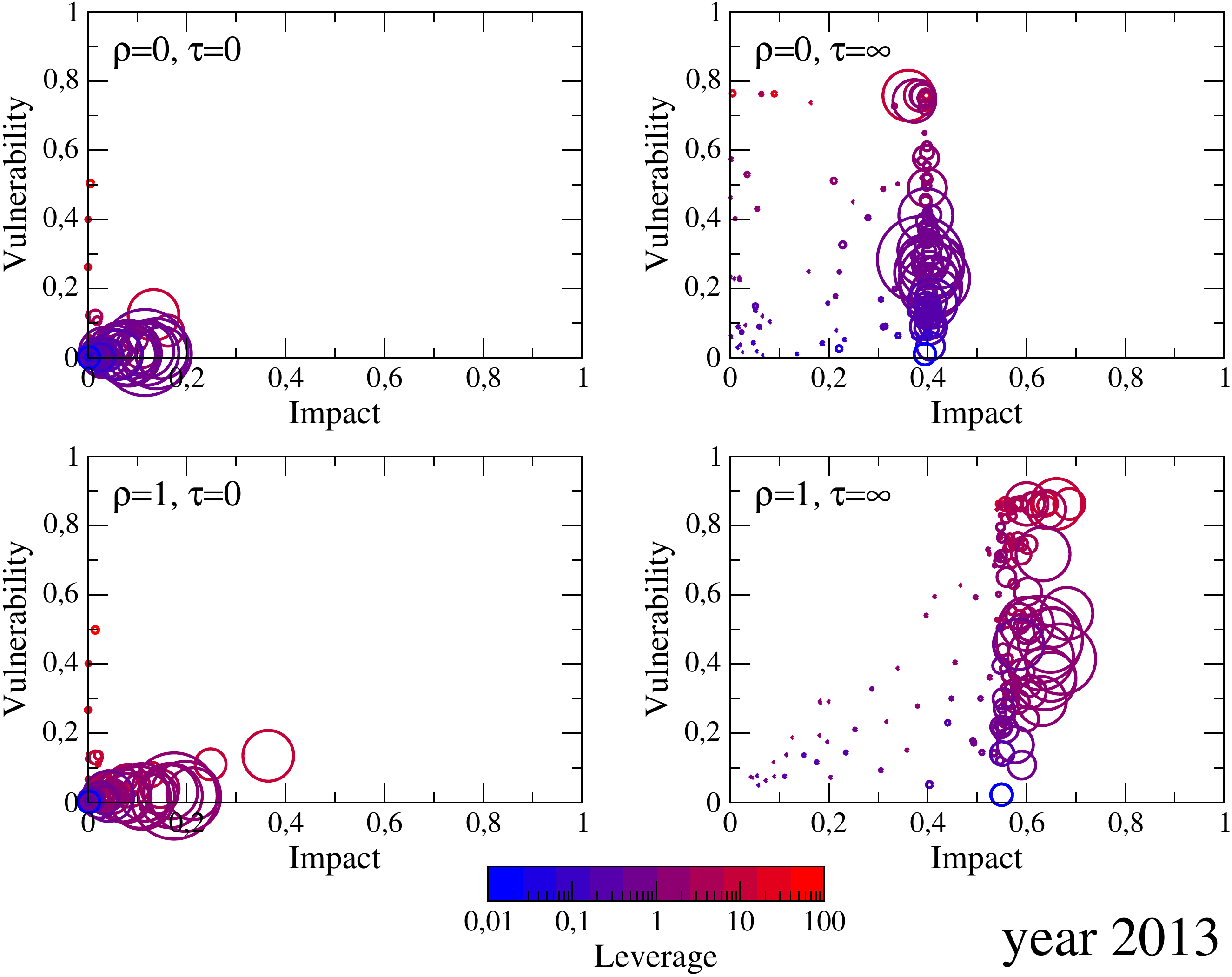}
\end{center}
\end{figure}

\end{document}